\documentclass[twocolumn]{aastex7}
\usepackage[T1]{fontenc}
\usepackage{amsmath}
\usepackage{CJKutf8}

\graphicspath{{./fig/}}

\begin{document}

\title{Formation of Multiple Dynamical Classes in the Kuiper Belt via Disk Dissipation}

\begin{CJK*}{UTF8}{bsmi}
\author[orcid=0000-0001-7433-1177,gname='Tommy Chi Ho',sname='Lau']{Tommy Chi Ho Lau (劉智昊)}
\affiliation{Department of Astronomy \& Astrophysics, University of Chicago, Chicago, IL 60637, USA}
\affiliation{University Observatory, Faculty of Physics, Ludwig-Maximilians-Universität München, Scheinerstr. 1, D-81679, Munich, Germany}
\email[show]{tommylauch@uchicago.edu}

\author[orcid=0000-0002-1899-8783,sname='Birnstiel']{Til Birnstiel}
\affiliation{University Observatory, Faculty of Physics, Ludwig-Maximilians-Universität München, Scheinerstr. 1, D-81679, Munich, Germany}
\affiliation{Exzellenzcluster ORIGINS, Boltzmannstr. 2, D-85748 Garching, Germany}
\email{til.birnstiel@lmu.de}

\author[orcid=0000-0002-1589-1796,sname='Stammler']{Sebastian M. Stammler}
\affiliation{University Observatory, Faculty of Physics, Ludwig-Maximilians-Universität München, Scheinerstr. 1, D-81679, Munich, Germany}
\email{stammler@usm.lmu.de}

\author[orcid=0000-0002-9128-0305,sname='Dr{\k{a}}{\.z}kowska']{Joanna Dr{\k{a}}{\.z}kowska}
\affiliation{Max Planck Institute for Solar System Research, Justus-von-Liebig-Weg 3, D-37077 Göttingen, Germany}
\email{drazkowska@mps.mpg.de}

\begin{abstract}
Planetesimal formation likely lasted for millions of years in the solar nebula, and the cold classicals in the Kuiper Belt are suggested to be the direct products of streaming instability. The presence of minor planetary bodies in the outer solar system and the exo-Kuiper belts provide key constraints to planet formation models. In this work, we connected dust drift and coagulation, planetesimal formation, N-body gravity, pebble accretion, planet migration, planetary core accretion, gap opening, and internal photoevaporation in one modeling framework. We demonstrate that multiple classes of minor planets, or planetesimals, can form during disk dissipation and remain afterwards, including a scattered group, a resonant group, and a dynamically cold group. Significant growth by pebble accretion was prevented by both dynamical heating due to the giant planet in the system and rapid dispersal of the disk toward the end of its lifetime. We also conducted a parameter study which showed that this is not a universal case, where the outcome is determined by the competition for dust between planetesimal formation and pebble accretion. Combining this scenario with sequential planet formation, this model provides a promising pathway toward an outer solar system formation model.
\end{abstract}


\section{Introduction} \label{sec:intro}
The timing and location of planetesimal formation are critical to the final architecture of a planetary system. In particular, the solar system's minor bodies are one of the key constraints of the classic solar system models (\citealp[see][for review]{Nesvorny2018}), including the Nice model \citep[e.g.][]{Morbidelli2005,Tsiganis2005} and the early instability model \citep[e.g.][]{Clement2017,Deienno2018,Liu2022}. Meanwhile, current N-body planet formation models \citep[e.g.][]{Matsumura2017,Bitsch2019,Matsumura2021,Lau2024} generally assume an initial distribution of planetesimals or embryos and demonstrate difficulties in explaining the formation of the solar system's giant planets. Furthermore, meteoritic records (\citealp[see][for review]{Kleine2020}) show that planet formation spans millions of years in the solar system. This emphasizes the need for a global planet formation that includes dust evolution and planetesimal formation in the protoplanetary disk.

Recently, streaming instability \citep{Youdin2005,Johansen2007} has become the prevailing mechanism to form planetesimals and \cite{Squire2020} further provided a physical picture of the process. First, a relatively dense inward-drifting dust clump is required. When the clump drags the gas around it as a result of back reaction, the gas also deflects azimuthally in the direction of disk rotation due to the Coriolis force. This motion of the gas pushes the dust along, causing an outward motion relative to the overall drift. The dust density is further enhanced and a feedback loop is created. Evidently, a high dust-to-gas ratio and a large Stokes number of the dust particles are required to trigger streaming instability (\citealp[e.g.][]{Carrera2015,Li2021,Yang2017,Lim2024}; \citealp[see][for review]{Simon2024}). \cite{Nesvorny2019} further showed that streaming instability can also produce a distribution of binary orientations that is in good agreement with that of the cold classical binaries observed in the Kuiper Belt.

Among the different mechanisms that can create disk substructure facilitating planetesimal formation, photoevaporation is likely a promising candidate. The cavity, which opens toward the end of the disk's lifetime, could trap dust and trigger planetesimal formation while further growth could be prevented, even for the dynamically cold ones. \cite{Carrera2017,Ercolano2017} first demonstrated the scenario in which photoevaporation preferentially removes gas and the dust-to-gas ratio is enhanced, making it possible to fulfill the streaming instability conditions. The cavity that opens during the transition stage of a photoevaporating disk further retains dust from being lost through disk accretion. However, the subsequent evolution of the formed planetesimals was not studied, which requires follow-up investigations. Moreover, the presence of amorphous ice suggested by observations of comets implies a low abundance of radioactive isotopes, which further supports the late formation of these bodies (\citealp[e.g.][]{Prialnik1987,Mousis2017}; \citealp[see][for review]{Guilbert-Lepoutre2024}).

In a protoplanetary disk, the timing and location of planetesimal formation have become an active research topic \citep[e.g.][]{Drazkowska2016,Carrera2017,Schoonenberg2018,Lenz2019,Lenz2020}. Multiple dynamical classes of the Kuiper Belt objects, including the cold classical objects, resonant objects and the scattered disk, as well as the exo-Kuiper belts observed, such as \cite{MacGregor2017, Marino2017} (\citealp[see][for review]{Marino2022}), further constrain the formation mechanism and dynamical history of these systems.

In this work, we further develop the global planet formation model presented in \cite{Lau2022,Lau2024b} to include the effect of internal photoevaporation and model planetesimal formation and evolution in the final stage of a protoplanetary disk. We identified a scenario that leads to the formation of multiple dynamical classes of small planetary bodies surviving after disk dissipation. In the following, Sec. \ref{sec:methods} summarizes the methods adopted in \cite{Lau2022,Lau2024b} and the new implementations in this work. The results are presented in Sect. \ref{sec:results}, followed by the discussions and outlooks in Sect. \ref{sec:dis}. Our findings are summarized in Sect. \ref{sec:concl}.

\section{Methods} \label{sec:methods}
We employed the dust and gas evolution code DustPy v1.0.8 \citep{Stammler2022} and the symplectic N-body integrator SyMBAp v1.8 \citep{Lau2023}, which is a parallelized version of the Symplectic Massive Body Algorithm (SyMBA; \citealp{Duncan1998}). The coupling of the two codes to construct a planet formation model was presented in \cite{Lau2022,Lau2024b} but disk dissipation had not been previously included. In this work, we added internal photoevaporation to model the final stage of planet formation and the complete lifetime of a protoplanetary disk. The following summarizes the retained methods in \cite{Lau2022,Lau2024b} and describes the modifications and new implementations in detail.

\subsection{Disk model} \label{sec:m:disk}
We considered an axis-symmetric protoplanetary disk around a solar-type star. DustPy simulates the viscous evolution of the gas, the coagulation, fragmentation, advection, as well as diffusion of the dust. The following describes the different parts of the disk model.

\subsubsection{Gas component} \label{sec:m:gas}
The disk was assumed to be in vertical hydrostatic equilibrium and, radially, the gas surface density $\Sigma_\mathrm{g}$ evolved in time $t$ according to the advection-diffusion equation \citep{Lust1952,Lynden-Bell1974} with additional terms to include the angular momentum injection \citep{Lin1986,Trilling1998} and the mass loss due to internal photoevaporation. The equation reads
\begin{equation}\label{eq:adv-diff}
	\frac{\partial \Sigma_\mathrm{g}}{\partial t}=\frac{3}{r}\frac{\partial}{\partial r}\left[ r^{1/2} \frac{\partial}{\partial r}(\nu \Sigma_\mathrm{g} r^{1/2}) -\frac{2\Lambda\Sigma_\mathrm{g}}{3\Omega_\mathrm{K}}\right] + \dot{\Sigma}_\mathrm{g,IPE},
\end{equation}
where $r$ is the distance from the star, $\nu$ is the kinematic viscosity, $\Lambda$ is the profile of the specific angular momentum injection rate, and $\dot{\Sigma}_\mathrm{g,IPE}$ is the rate of change of surface density due to internal photoevaporation. The local Keplerian orbital frequency is given by $\Omega_\mathrm{K}=\sqrt{GM_\odot/r^{3}}$ with the gravitational constant $G$. The back-reaction from the dust is neglected in this work.

The initial gas surface density $\Sigma_{\mathrm{g}}(t=0)$ is given by
\begin{equation}
	\Sigma_{\mathrm{g}}(t=0) = \frac{M_\mathrm{disk}}{2\pi r_\mathrm{c}^2}\left( \frac{r}{r_\mathrm{c}}\right) ^{-1}\exp\left(-\frac{r}{r_\mathrm{c}}\right),
\end{equation}
with the initial mass of the disk $M_\mathrm{disk}$, and the characteristic radius $r_\mathrm{c}$. We set $M_\mathrm{disk}=\{0.025,0.05\}M_\odot$ and $r_\mathrm{c}=50\ \mathrm{au}$, which imply $\Sigma_{\mathrm{g}} \ (t=0,r=5\ \mathrm{au})\approx \{134.6, 269.2\}\ \mathrm{g}\ \mathrm{cm}^{-2}$.

The \cite{Shakura1973} $\alpha$-parametrization was adopted for $\nu$ such that
\begin{equation}
	\nu = \alpha c_\mathrm{s} H_\mathrm{g},
\end{equation}
with the speed of sound $c_\mathrm{s}$ and the disk scale height $H_\mathrm{g}\equiv c_\mathrm{s}/\Omega_\mathrm{K}$. The viscosity parameter $\alpha=5\times10^{-4}$ was set in this work. The isothermal sound speed was used and given by $c_\mathrm{s} = \sqrt{k_\mathrm{B}T/\mu}$ with the Boltzmann constant $k_\mathrm{B}$, the midplane temperature $T$ and the mean molecular weight of the gas $\mu=2.3m_\mathrm{p}$. The disk was assumed to be passively irradiated by the solar luminosity at a constant angle of 0.05, which gave the midplane temperature profile
\begin{equation}
	T\approx221 \left( \frac{r}{\mathrm{au}}\right) ^{-1/2} \mathrm{K}.
\end{equation}
This setup yielded the dimensionless gas disk scale height
\begin{equation}
	\hat{h}_\mathrm{g}\equiv\frac{H_\mathrm{g}}{r}\approx0.0299\left( \frac{r}{\mathrm{au}}\right)^{1/4}.
\end{equation}
And, the volumetric gas density was given by 
\begin{equation}
	\rho_\mathrm{g}=\rho_\mathrm{g}(z=0)\exp \left(-\frac{z^2}{2H_\mathrm{g}^2} \right)
\end{equation}
with the gas density at midplane
\begin{equation}
	\rho_\mathrm{g}(z=0)=\frac{1}{\sqrt{2\pi}}\frac{\Sigma_\mathrm{g}}{H_\mathrm{g}}.
\end{equation}
The midplane pressure gradient parameter $\eta$ is then given by
\begin{equation}
	\eta=-\frac{\hat{h}_\mathrm{g}^2}{2}\frac{\partial\ln P}{\partial\ln r}, \label{eq:eta}
\end{equation}
with the midplane gas pressure $P$. A logarithmic radial grid was adopted with 260 cells from 1 to 1000 au, including 200 cells from 2 to 50 au to refine the grid of the interested region.

\subsubsection{Torque deposition} \label{sec:m:torq}
In Eq.~(\ref{eq:adv-diff}), we implemented the torque deposition term from \cite{Lin1986} and \cite{Trilling1998} for the initial gap and planetary gap. This is in contrast to the implementation in \cite{Lau2022,Lau2024b}, where the initial pressure bump and planetary gap were imposed on the gas by modifying the $\alpha$-parameter. This change is due to the fact that the disk is no longer in steady-state accretion during dissipation and the previous implementation resulted in an unphysical retention of gas at where the viscosity is increased through the $\alpha$-parameter in our test simulations \citep{Lau2024b}.

As noted in \cite{Lau2024b}, other works \citep[e.g.][]{Lin1986,Armitage2002,DAngelo2010} provided profiles of the torque density exerted by a planet. However, applying the profile by \cite{DAngelo2010} did not produce consistent results in our one-dimensional model with respect to other works on planetary gap profile \citep[e.g.][]{Kanagawa2015,Duffell2020}. Therefore, an equivalent $\Lambda$ is derived by assuming a steady-state accretion with the target surface density profile. This assumption is inline with \cite{Duffell2020}, where their gap profile has been applied (Sect. \ref{sec:m:gap}), and they stated that the planets are effectively in a steady-state disk in their hydrodynamical simulations, which is also the case in other studies on planetary gap profile \citep[e.g.][]{Kanagawa2015,Duffell2015}. Since this $\Lambda$ profile is specific to the planet, it remains unchanged when the disk starts dissipating and departs from the steady state. The derivation is first described in \cite{Lau2024a} and detailed in the following.

Neglecting the photoevaporation term, eq. (\ref{eq:adv-diff}) can be rewritten as
\begin{equation}
	\frac{\partial \Sigma_\mathrm{g}}{\partial t}=-\frac{1}{r}\frac{\partial}{\partial r}\left[ r \Sigma_\mathrm{g}( v_\nu+v_\Lambda) \right]
\end{equation}
with the accretion velocity
\begin{equation}
	v_\nu=-\frac{3\nu}{r}\frac{\partial}{\partial \log r} \log \left( \nu \Sigma_\mathrm{g} r^{1/2} \right)\label{eq:visc_vel}
\end{equation}
and the additional velocity due to torque injection for a specific substructure
\begin{equation}
	v_\Lambda=\frac{2\Lambda'}{r\Omega_\mathrm{K}}.
\end{equation}
Here, $\Lambda'$ corresponds to a specific substructure and the total contributions from all substructures are summed together as described later in eq. (\ref{eq:lambda}). As a part of the steady-state assumption stated above, disk accretion is impeded by the substructure, which can be expressed as
\begin{equation}\label{eq:steady}
	\Sigma_\mathrm{g,eq} (v_{\nu,\mathrm{eq}}+v_\Lambda)=\Sigma_\mathrm{g,0}v_{\nu,0},
\end{equation}
where $\Sigma_\mathrm{g,eq}$ and $v_{\nu,\mathrm{eq}}$ are the gas surface density and accretion velocity at the steady state with the imposed gap respectively. Similarly, $\Sigma_\mathrm{g,0}$ and $v_{\nu,0}$ are the unperturbed gas surface density and the unperturbed accretion velocity respectively. Assuming a constant disk accretion rate, which is given by $\dot{M}_\mathrm{disk}=3\pi\nu\Sigma_{\mathrm{g,0}}$ \citep{Lynden-Bell1974}, yields
\begin{equation}\label{eq:steady_acc}
	\frac{\partial(\Sigma_\mathrm{g,0}\nu)}{\partial r}=0.
\end{equation}
And, we define the target surface density profile by
\begin{equation}\label{eq:f_r}
	f(r) \equiv \frac{\Sigma_\mathrm{g,eq}}{\Sigma_\mathrm{g,0}}.
\end{equation}

Substituting eq. (\ref{eq:steady_acc}) \& (\ref{eq:f_r}) into eq. (\ref{eq:visc_vel}), the accretion velocity with the imposed gap and the unperturbed one are, respectively, given by
\begin{equation}
	v_{\nu,\mathrm{eq}}=-\frac{3\nu}{r}\left(\frac{\partial\log f}{\partial \log r} +\frac{1}{2}\right),
\end{equation}
and
\begin{equation}
	v_{\nu,0}=-\frac{3\nu}{2r}.
\end{equation}
With eq. (\ref{eq:visc_vel}) to (\ref{eq:steady}), the angular momentum injection rate profile corresponds to this specific substructure is 
\begin{equation}
	\Lambda'(f)=-\frac{3\nu\Omega_\mathrm{K}}{2}\left(\frac{1}{2f}-\frac{\partial\log f}{\partial \log r} - \frac{1}{2}\right).
\end{equation}
The respective contributions were summed together when there were multiple substructures in the disk. In other words, the total injection rate is given by
\begin{equation}\label{eq:lambda}
	\Lambda=\Lambda'(f_\mathrm{bump})+ \sum_{i} \Lambda'(f_{\mathrm{gap},i})
\end{equation}
for all planet $i$ and is applied in eq (\ref{eq:adv-diff}). The target surface density profile due to the initial pressure bump $f_\mathrm{bump}$ is described in Sect. \ref{sec:m:bump} and that due to a planet $f_{\mathrm{gap},i}$ is described in Sect. \ref{sec:m:gap}. This torque was exclusively applied to the gas while the dust experienced an indirect effect due to the change in the gas velocity. 

\subsubsection{Internal photoevaporation} \label{sec:m:PE}
We followed \cite{Picogna2019} to calculate the gas loss rate of a disk that is under the X-ray and extreme ultraviolet irradiation from the central star. Generally, the radial profile of a primordial disk is believed to be dominated by viscous evolution. As the accretion rate falls and becomes comparable to the loss rate due to photoevaporation, a cavity opens and the disk enters the transition stage. After the inner disk is depleted, the outer edge of the cavity is directly irradiated by the star and the rate of photoevaporation is locally enhanced. \cite{Garate2021} first applied this prescription in DustPy and the following summarizes the implementation. In this work, we have neglected the effect of far-ultraviolet radiation \citep{Gorti2015,Carrera2017} due to its complexity because heating depends on the abundances of small grains and polycyclic aromatic hydrocarbon molecules. This will likely truncate the outer region early in the disk's lifetime.

In the case of a primordial disk, the total mass loss rate $-\dot{M}_\mathrm{g,pr}$ is given by
\begin{equation}
	\begin{split}
	&\log\left( \frac{-\dot{M}_\mathrm{g,pr}}{M_\odot\mathrm{\ yr}^{-1}} \right) =\\
	&A_\mathrm{L}\exp\left\{ \frac{[\ln\left( \log L_\mathrm{X}\right)-B_\mathrm{L}]^2}{C_\mathrm{L}} \right\} +D_\mathrm{L}
	\end{split}
\end{equation}
with the X-ray luminosity of the star $L_\mathrm{X}$. The fitting parameters are $A_\mathrm{L}=-2.7326$, $B_\mathrm{L}=3.3307$, $C_\mathrm{L}=-2.9868\times 10^{-3}$ and $D_\mathrm{L}=-7.2580$. The loss rate of surface density in this case is
\begin{equation}
	\begin{split}
	&-\dot{\Sigma}_\mathrm{g,pr} = \\
	&\frac{\dot{M}_\mathrm{r}}{2\pi r_\mathrm{au}^2} M_\odot \mathrm{au}^{-2}\mathrm{yr}^{-1} \times \sum_{k=1}^{6} ka_k\log^{k-1}r_\mathrm{au},
	\end{split}
\end{equation}
where the radial mass loss factor is
\begin{equation}
	\dot{M}_\mathrm{r} = \dot{M}_\mathrm{g,pr}10^n,
\end{equation}
with the index
\begin{equation}
	n=\sum_{k=0}^{6} a_k\log^k r_\mathrm{au}
\end{equation}
and $r_\mathrm{au}\equiv r/\mathrm{au}$. The fitting parameters are
$a_0=-2.8562$, $a_1=5.7248$, $a_2=-11.4721$, $a_3=16.3587$, $a_4=-12.1214$, $a_5=4.3130$ and $a_6=-0.5885$.

In the case of a transition disk, the loss rate of surface density is
\begin{equation}
	-\dot{\Sigma}_\mathrm{g,tr} = \frac{b_0b_1^xx^{b_2-1}\left(x\ln b_1+b_2 \right)}{r_\mathrm{au}} M_\odot \mathrm{au}^{-2}\mathrm{yr}^{-1}
\end{equation}
with the fitting parameters $b_0=0.11843$, $b_1=0.99695$ and $b_2=0.48835$. And, $x$ is the distance from the radial location of the outer edge of the cavity $r_\mathrm{cavity}$, which is defined by
\begin{equation}
	x\equiv \frac{r-r_\mathrm{cavity}}{\mathrm{au}}.
\end{equation}
We followed the implementation by \cite{Garate2021} that the cavity was considered open when $r_\mathrm{cavity}\geq 7.5 \mathrm{au}$ since the inner disk is likely to have been fully removed. The general loss rate of the surface density is
\begin{equation}
	\dot{\Sigma}_\mathrm{g,IPE} = \begin{cases}
		\dot{\Sigma}_\mathrm{g,pr} &r_\mathrm{cavity} < 7.5\mathrm{au}\\
		\dot{\Sigma}_\mathrm{g,tr} &r_\mathrm{cavity}\geq 7.5\mathrm{au}.
	\end{cases}
\end{equation}
Dust entrainment in the photoevaporation wind was also considered and is described in Sect. \ref{sec:m:dust_PE}.

The values considered for the X-ray luminosity of the star $L_\mathrm{X}$ are $\{2\times10^{29},5\times10^{29},10^{30}\}\mathrm{\ erg\ s}^{-1}$ in this work. The chosen values are at the high end of the measured distribution of T Tauri stars \citep[e.g.][]{Preibisch2005}. This is due to the computation limit that short-lived disks are more affordable to model and this particular work focuses on the final stage of planet formation during disk dissipation. The disk was considered to have dispersed when the gas surface density was less than $10^{-3} \mathrm{g}\ \mathrm{cm}^{-2}$ for all cells.

\subsubsection{Dust component} \label{sec:m:dust}
The dust surface density $\Sigma_{\mathrm{d}}$ is initially given by
\begin{equation}
	\Sigma_{\mathrm{d}} (t=0) = Z\Sigma_{\mathrm{g}}(t=0)
\end{equation}
with the global dust-to-gas ratio $Z$ set at the solar metallicity of 0.01. We followed \cite{Mathis1977}, also known as the MRN size distribution of the interstellar medium, for the initial size distribution of the dust grains. The maximum initial size was set at $1\ \mu\mathrm{m}$ with the internal density of $1.67\ \mathrm{g}\ \mathrm{cm}^{-3}$ assumed. A total of 141 dust mass bins logarithmically spaced from $10^{-12}$ to $10^8$ g were used. Each dust mass species was evolved through transportation with the advection-diffusion equation \citep{Clarke1988} coupled to growth and fragmentation with the Smoluchowski equation. The fragmentation velocity was set to $5\ \mathrm{m}\ \mathrm{s}^{-1}$, which is close to the value found in \cite{Wurm2005,Wada2013} for icy dust particles. Nonetheless, we note that the fragmentation velocity should depends on grain property in reality, including size and composition, while condensation lines are currently not included in our model. The water ice line will likely have the greatest impact when the model is extended closer to the star.

For collision velocities above this value, the dust particles were assumed to fragment. The Stokes number St$_i$ corresponding to dust species $i$ was calculated by considering the Epstein and the Stokes I regimes. The dust scale height of each dust species $H_{\mathrm{d},i}$ was calculated according to \cite{Dubrulle1995},
\begin{equation}
	H_{\mathrm{d},i}=H_\mathrm{g}\sqrt{\frac{\alpha}{\alpha+\mathrm{St}_i}}, \label{eq:h_d}
\end{equation}
assuming $\mathrm{St}_i<1$. And, the volumetric dust density of species $i$ is given by 
\begin{equation}
	\rho_{\mathrm{d},i}=\rho_{\mathrm{d},i}(z=0) \exp \left(-\frac{z^2}{2H_{\mathrm{d},i}^2}\right) 
\end{equation}
with the corresponding dust density at midplane
\begin{equation}
	\rho_{\mathrm{d},i}(z=0)=\frac{1}{\sqrt{2\pi}}\frac{\Sigma_{\mathrm{d},i}}{H_{\mathrm{d},i}}.
\end{equation}
Further details of the algorithms for the disk model are described in \cite{Stammler2022}.

\subsubsection{Dust entrainment} \label{sec:m:dust_PE}
Generally, the small dust high above the midplane is well-coupled to the photoevaporative wind and is removed. \cite{Garate2021} also implemented the loss due to dust entrained in the photoevaporative winds based on the 2-D model by \cite{Franz2020}. We followed their implementation and applied this process to dust with radius $a$ smaller than $10\,\mu\mathrm{m}$ and above $3H_\mathrm{g}$. In other words, the rate of loss of dust surface density for dust species $i$ is

\begin{equation}
	-\dot{\Sigma}_{\mathrm{d,IPE},i} = \begin{cases}
		\epsilon_{\mathrm{ent},i}\dot{\Sigma}_\mathrm{g,IPE} &a_i<10 \ \mu\mathrm{m}\\
		0 &a_i\geq 10 \ \mu\mathrm{m}
	\end{cases}
\end{equation}
and the corresponding dust-to-gas ratio is
\begin{equation}
	\epsilon_{\mathrm{ent},i} =\frac{ \int_{3H_\mathrm{g}}^{\infty} \rho_{\mathrm{d},i} \ \mathrm{d}z}{\int_{3H_\mathrm{g}}^{\infty} \rho_\mathrm{g} \ \mathrm{d}z}.
\end{equation}

As the gas surface density decreases, the dust fragmentation limit decreases as well in terms of particle size. However, the finite range of the dust mass bins cannot describe an arbitrarily small dust mass distribution. This can result in a numerical issue where an unphysical amount of dust remains in the lowest mass bin because it cannot further fragment. The issue deteriorates when the lowest mass bin also has a high Stokes number due to a low gas surface density, which means the dust cannot be entrained and removed through our implementation of photoevaporative wind. Therefore, in a radial cell, if more than 10\% of the dust mass is in the smallest bin and also if this bin has a Stokes number that is greater than $10^{-4}$, we assumed the dust is well mixed with the gas when evaluating dust entrainment. This implementation ensures that the dust is removed with the gas when the gas surface density is close to the floor value.

\subsubsection{Initial pressure bump} \label{sec:m:bump}
In a smooth protoplanetary disk, the condition for the streaming instability is generally difficult to reach because dust drifts without interruption and does not concentrate. A pressure bump also creates a favorable environment for the subsequent growth of planetesimals due to enhanced pebble density and reduced radial drift velocity \citep{Lau2022}. To trigger planetesimal formation and planet formation, an initial pressure bump was introduced to the disk following the model by \cite{Dullemond2018a}. Although we do not investigate the cause of such bump in this work, possible nonplanetary causes include sublimation \citep{Saito2011}, instabilities \citep{Takahashi2014,Flock2015,Dullemond2018}, late-stage infall \citep{Gupta2023}, and the edge of the magnetorotational-instability dead zone \citep{Pinilla2016}. The change relative to the unperturbed surface density is given by
\begin{equation}\label{eq:f_bump}
	\begin{split}
	f_\mathrm{bump}(r) &\equiv \frac{\Sigma_\mathrm{g,bump}}{\Sigma_\mathrm{g,0}}\\
	&=\exp\left[ -A \exp\left(-\frac{(r-r_0)^2}{2w^2} \right) \right]
	\end{split}
\end{equation}
with the unperturbed gas surface density $\Sigma_\mathrm{g,0}$, the gas surface density with the bump $\Sigma_\mathrm{g,bump}$, the amplitude $A=2$, the location $r_0=6$ au, and the width $w=2H_\mathrm{g}(r=r_0)$. These parameters are chosen based on the location of Jupiter and test runs which show that such a combination can trigger planetesimal formation. Nonetheless, further parameter studies are required for future works. The initial pressure bump is then applied through torque deposition as described in Sect. \ref{sec:m:torq}.

We note that the Rossby wave instability (RWI) may occur in a narrow pressure bump, which can disrupt the pressure bump itself. We have applied the criterion by \cite{Chang2023} and it shows that the above setup is RWI-stable.

\subsubsection{Planetesimal formation} \label{sec:m:plts_form}
Following \cite{Lau2024b}, we adopted the Toomre-like instability parameter $Q_\mathrm{p}$ for the gravitational collapse of the streaming instability-induced dust filaments and the initial planetesimal mass function \citep{Gerbig2023}. Meanwhile, multiple works (\citealp[e.g.][]{Carrera2015,Li2021,Yang2017,Lim2024}; \citealp[see][for review]{Simon2024}) showed that large particles are also required to trigger clumping to form such dense dust filaments. Therefore, we only considered dust with the Stokes number of at least $10^{-3}$ when evaluating planetesimal formation as detailed below.

The instability criterion is $Q_\mathrm{p}\leq1$, with
\begin{equation}
	Q_\mathrm{p}=\sqrt{\frac{\delta}{\mathrm{St}_\mathrm{avg}}}\frac{c_\mathrm{s}\Omega_\mathrm{K}}{\pi G\Sigma_\mathrm{d, local}},
\end{equation}
where $\mathrm{St}_\mathrm{avg}$ is the mass-averaged Stokes number of the dust with a Stokes number of at least $10^{-3}$ in the cell. The local dust surface density in the filament $\Sigma_\mathrm{d, local}$ was assumed to be $10\times\Sigma_\mathrm{d}(\mathrm{St}\geq10^{-3})$. This is motivated by streaming instability simulations \citep[e.g.][]{Simon2016,Schafer2017}, where the filaments are an order of magnitude more dense than the averaged dust density prior to gravitational collapse. The small-scale diffusion parameter $\delta$ was set at $5\times10^{-6}$, which is motivated by the streaming instability simulations in \cite{Schreiber2018}.

Dust was converted into planetesimals based on the treatments by \cite{Drazkowska2016} and \cite{Schoonenberg2018}. The criterion of $Q_\mathrm{p}\leq1$ was combined with the smooth planetesimal formation activation function from \cite{Miller2021}, which is given by
\begin{equation}
	\mathcal{P_\mathrm{pf}}= \frac{1}{1+\exp{[10\times(Q_\mathrm{p}-0.75)]}}, \label{eq:Ppf}
\end{equation}
and evaluated at each radial grid cell. If any cell also satisfied the criterion of $\rho_\mathrm{d}(\mathrm{St}\geq10^{-3})/\rho_\mathrm{g}\ge1$ in the midplane, the dust surface density for each dust species $i$ with $\mathrm{St}\geq10^{-3}$ was reduced by
\begin{equation}
	\frac{\partial \Sigma_{\mathrm{d},i}}{\partial t}=-\mathcal{P_\mathrm{pf}}\Sigma_{\mathrm{d},i}\frac{\zeta}{t_{\mathrm{set},i}}.
\end{equation}
The planetesimal formation efficiency per settling time was $\zeta=10^{-3}$ and the settling time of dust species $i$ is defined by $t_{\mathrm{set},i}\equiv1/(\mathrm{St}_i\Omega_\mathrm{K})$. \cite{Simon2016} shows that about half of the pebble being transformed to planetesimals in tens of the orbital timescales. Although this translates to a $\zeta$ of about $10^{-2}$, \cite{Drazkowska2016} note that these models correspond to a specific Stokes number of the dust and the value of $\zeta$ would be lower for smaller dust. Since $\zeta$ is poorly constrained, we opted for a conservative value compared to other studies \citep[e.g.][]{Schoonenberg2018,Miller2021}. The effect of this uncertainty are discussed further in Sect. \ref{dis:SI_PA}.

The removed dust was then summed over all dust species and added to the radial profile of planetesimal mass surface density. To realize the planetesimals as $N$-body particles, we first drew the location of the new planetesimal using this density profile as a probability function. Then, we drew the mass of this planetesimal according to the initial mass function given by \cite{Gerbig2023}, which results from the stability analysis of the dispersion relation for dust under the influence of turbulent diffusion \citep{Klahr2021}.
We refer the readers to eq. (20) in \cite{Gerbig2023} for the exact expression of the probability density function.

As described in \cite{Lau2022}, the eccentricity $e$ and the inclination $i$ in radians were, respectively, drawn from two Rayleigh distributions with the scale parameters of $10^{-6}$ and $5\times10^{-7}$. The means of $e$ and $i$ are given by multiplying $\sqrt{\pi/2}$ to the respective scale parameter. The rest of the angular orbital elements in radians were drawn uniformly from 0 to $2\pi$. The physical radius $R_\mathrm{p}$ was calculated by assuming that the internal density $\rho_\cdot = 1.5\,\mathrm{g}\,\mathrm{cm}^{-3}$. The drawn planetesimal mass was then subtracted from the surface density of the nearest radial grid cells. The realization stopped when the total remaining mass was less than the latest drawn mass that had not been introduced as a new particle. To avoid bias toward lower mass, the residue of the planetesimal mass surface density and the last drawn planetesimal mass were retained for the next time step. In the fiducial case introduced later in Sect. \ref{sec:m:num}, about 6,000 to 8,000 planetesimals were formed throughout each run.

\subsection{Planetesimal evolution} \label{sec:m:plts_evol}
SyMBAp was employed to integrate the direct $N$-body gravitational interactions among the central star and the planetesimals. Collisions were assumed to be perfectly inelastic and mass was accreted onto the more massive particle. The effects of pebble accretion (Sect. \ref{sec:m:PA}), gas accretion (Sect. \ref{sec:m:gas_acc}) and planetary gap opening (Sect. \ref{sec:m:gap}), gas drag, and planet-disk interactions (Sect. \ref{sec:main:mig} \& \ref{sec:m:mig}) were also included to evolve the planetesimals with feedback to the disk, which are further described in the following.

\subsubsection{Pebble accretion} \label{sec:m:PA}
The implementation of pebble accretion was identical to that presented in \cite{Lau2022} and is summarized below. The pebble mass flux corresponding to dust species $i$ at $r$ was given by
\begin{equation}
	\dot{M}_{\mathrm{peb},i}=2\pi rv_{\mathrm{drift},i} \Sigma_{\mathrm{d},i},
\end{equation}
where the corresponding pebble drift speed is $v_{\mathrm{drift},i}=2\mathrm{St}_i|\eta|r\Omega_\mathrm{K}$ \citep{Weidenschilling1977}.
Then, the pebble accretion efficiency factor $\epsilon_{\mathrm{PA},i}$ by \cite{Liu2018} and \cite{Ormel2018} was evaluated for each $N$-body particle, which can be a planetesimal or a planet, for all dust species $i$. The pebble accretion rate was then given by summing the contributions from all dust species, which is
\begin{equation}
	\dot{m}_\mathrm{pa}=\sum_i \epsilon_{\mathrm{PA},i} \dot{M}_{\mathrm{peb},i}.
\end{equation}
The accreted pebble mass was then subtracted from the respective dust species and the local radial cell at the next and immediate communication step. Since dust and gas evolved consistently in this model, the gap opened by a planet (Sect. \ref{sec:m:gap}) shall interrupt the pebble flux resulting in pebble isolation. We note that trapped pebbles fragment into  small dust, which can leak through the planetary gap, especially when it is shallow \citep[e.g.][]{Stammler2023}. Nonetheless, pebble accretion is much less efficient for small dust as in the adopted prescription. The rate of pebble accretion by a core gradually decreases as it grows and carves a wider and deeper gap that becomes more effective in trapping dust. The effect of pebble accretion on the gas accretion rate is described below in Sect. \ref{sec:m:gas_acc} and explicitly included in eq. (\ref{eq:gas_acc}), where gas accretion is only possible when the pebble accretion rate is much smaller than the potential gas accretion rate.

\subsubsection{Gas accretion} \label{sec:m:gas_acc}
We generally implemented the core accretion model \citep{Bodenheimer1986,Mizuno1980,Pollack1996} to account for gas accretion on a planetary core. The core is first in the thermal contraction phase where the hydrostatic equilibrium is maintained until it reaches a critical mass. Then, the runaway gas accretion phase starts. As noted in \cite{Lau2024b}, planetary gas accretion is still an active field of research \citep[e.g.][]{Szulagyi2016,Lambrechts2019,Schulik2019,Ormel2021,Brouwers2021} and a key uncertainty is the envelope opacity. To better understand the consequence, we opted to treat the envelope opacity as a free parameter, which also encapsulated the overall gas accretion efficiency. Furthermore, the models by \cite{Bitsch2015} and \cite{Chambers2021} only considered the limit imposed by cooling in the runaway accretion phase. This is likely the cause of the rapid growth in this phase as noted in \cite{Lau2024b}. Physically, the gas accretion should be limited by both the cooling rate of the envelope and the supply of gas in both the thermal contraction phase and the runaway accretion phase. In this work, we combined different prescriptions to construct a general accretion model as described in the following.

In the thermal contraction phase, we adopted the cooling-limited gas accretion rate by \cite{Bitsch2015} based on \cite{Piso2014} with the modification by \cite{Chambers2021} to account for the energy released by pebble accretion. This is given by
\begin{equation}\label{eq:gas_acc}
	\begin{split}
		&\dot{m}_\mathrm{cont}=\max\Bigg[ 0, \ 4.375\times10^{-9} \left(\frac{\kappa}{\mathrm{cm}^2\ \mathrm{g}^{-1}} \right)^{-1} \times \\
		&\left( \frac{\rho_\mathrm{c}}{5.5\ \mathrm{g}\ \mathrm{cm}^{-3}}\right)^{-1/6} \left( \frac{m_\mathrm{c}}{M_\oplus} \right)^{11/3}\left( \frac{m_\mathrm{env}}{M_\oplus} \right)^{-1} \times\\
		&\left( \frac{T}{81\ \mathrm{ K}}\right) ^{-1/2} M_\oplus \mathrm{yr}^{-1} - 15 \dot{m}_\mathrm{pa}\Bigg]
	\end{split}
\end{equation}
with the envelope opacity $\kappa=\{0.01,0.02\} \ \mathrm{cm}^2\ \mathrm{g}^{-1}$, the density of the core $\rho_\mathrm{c}=5.5\ \mathrm{g}\ \mathrm{cm}^{-3}$, the mass of the core $m_\mathrm{c}$, the mass of the envelope $m_\mathrm{env}$, the disk midplane temperature $T$ and the pebble accretion rate $\dot{m}_\mathrm{pa}$. Gas accretion transitioned to the runaway accretion phase when $m_\mathrm{env}= m_\mathrm{c}$.

In the runaway gas accretion phase, we followed \cite{Ikoma2000}, where the envelope collapses on the Kelvin-Helmholtz timescale, and the accretion rate was
\begin{equation}
	\dot{m}_\mathrm{run}=\frac{m}{\tau_\mathrm{KH}}
\end{equation}
with the Kelvin-Helmholtz contraction timescale
\begin{equation}
	\tau_\mathrm{KH}= 10^9 \left( \frac{m}{M_\oplus} \right) ^{-3} \left( \frac{\kappa}{1 \ \mathrm{cm}^{2} \ \mathrm{ g}^{-1} }\right) \mathrm{ yr},
\end{equation}
where the planet mass is $m$ and we followed \cite{Ida2004} for the indices.

To describe the gas supply, we applied the prescription by \cite{Tanigawa2002}, which is
\begin{equation}
	\dot{m}_\mathrm{flow}=0.29\Sigma_\mathrm{g}r^2\Omega_\mathrm{K} \left( \frac{m}{M_\odot}\right) ^{4/3}\hat{h}_\mathrm{g}^{-2}.
\end{equation}
Gas accretion was further restricted by the overall disk accretion rate as \cite{Lubow2006} suggested that the gas accretion rate is limited to about 80\% of that, which can be described by
\begin{equation}
	\dot{m}_\mathrm{acc}=0.8\dot{M}_\mathrm{disk}
\end{equation}
with the local disk accretion rate $\dot{M}_\mathrm{disk}$.

To summarize, the gas accretion rate of a planet was limited by the cooling rates of the corresponding phase, the gas supply and the disk accretion rate, i.e.
\begin{equation}
	\dot{m}_\mathrm{g} = \begin{cases}
		\min\left( \dot{m}_\mathrm{cont}, \dot{m}_\mathrm{flow}, \dot{m}_\mathrm{acc} \right)  &m_\mathrm{env}<m_\mathrm{c}\\
		\min\left( \dot{m}_\mathrm{run}, \dot{m}_\mathrm{flow}, \dot{m}_\mathrm{acc} \right)&m_\mathrm{env}\geq m_\mathrm{c}.
	\end{cases}
\end{equation}

\subsubsection{Physical radius} \label{sec:m:rad}
As a planetesimal or planet grew by many orders of magnitude in mass, the physical radius $R_\mathrm{p}$ was updated at each $N$-body timestep. For bodies with mass $m<0.1M_\oplus$, we assumed an internal density $\rho_\mathrm{s}$ of 1.5 $\mathrm{g}\ \mathrm{cm}^{-3}$ to calculate their radii, which is the same case for the newly formed planetesimals. For $0.1M_\oplus\leq m < 5M_\oplus $, we followed the mass-radius relationship of rocky planets given by \cite{Seager2007}, which is
\begin{equation}
	\begin{split}
	\log\left( \frac{R_\mathrm{p}}{3.3R_\oplus} \right) = -&0.209 + \frac{1}{3}\log\left( \frac{m}{5.5M_\oplus} \right)-\\
	&0.08\left( \frac{m}{5.5M_\oplus} \right)^{0.4}
	\end{split}
\end{equation}
with the radius of Earth $R_\oplus$.
For $m\geq 5M_\oplus$, we followed the mass-radius relationship applied in \cite{Matsumura2017}, which is
\begin{equation}
	R_\mathrm{p}=1.65\sqrt{\frac{m}{5M_\oplus}}R_\oplus.
\end{equation}

\subsubsection{Planetary gap opening} \label{sec:m:gap}
Generally, as a planet grows in mass and interacts with the disk gravitationally, a planetary gap starts to open. In our model, the nondimensional gap opening factor by \cite{Kanagawa2015} was evaluated for each planetary body, which is
\begin{equation}
	K=q^2\hat{h}_\mathrm{g}^{-5}\alpha^{-1}
\end{equation}
with the mass ratio $q\equiv m/M_\odot$. Only if the planet had a value of $K>0.25$, we further applied the empirical formula by \cite{Duffell2020} for the gap profile and evaluated its torque deposition as described in Sect. \ref{sec:m:torq}. The gap profile corresponding to a planet $i$ is given by
\begin{equation}\label{eq:duffell}
	f_{\mathrm{gap},i}(r) \equiv \frac{\Sigma_\mathrm{g, gap}}{\Sigma_{\mathrm{g},0}} = \left(1+\frac{0.45}{3\pi}\frac{\tilde{q}^2(r)}{\alpha\hat{h}_\mathrm{g}^5}\delta(\tilde{q}(r))\right)^{-1}
\end{equation}
with the gas surface density with the planetary gap $\Sigma_\mathrm{g, gap}$ and the value of $\hat{h}_\mathrm{g}$ evaluated at the planet's location. The radial profile function $\tilde{q}(r)$ is defined by
\begin{equation}
	\tilde{q}(r)\equiv\frac{q}{\left\lbrace 1+D^3\left[ (r/r_\mathrm{p})^{1/6}-1 \right] ^6 \right\rbrace^{1/3} }
\end{equation}
with the planet's radial distance from the star $r_\mathrm{p}$ and the scaling factor $D\equiv 7\hat{h}_\mathrm{g}^{-3/2}\alpha^{-1/4}$.
The function $\delta(x)$ is given by
\begin{equation}\label{eq:del_x}
	\delta(x) = \begin{cases}
		1+(x/q_w)^3, 	 &\mathrm{if}\ x<q_\mathrm{NL}\\
		\sqrt{q_\mathrm{NL}/x}+(x/q_w)^3, &\mathrm{if}\ x\geq q_\mathrm{NL}
	\end{cases}
\end{equation}
with the factor $q_\mathrm{NL}=1.04\hat{h}_\mathrm{g}^3$ and the factor $q_w=34 q_\mathrm{NL}\sqrt{\alpha/\hat{h}_\mathrm{g}}$.

\subsubsection{Gas drag and planet-disk interactions} \label{sec:main:mig}
All $N$-body particles experienced the combined effects of aerodynamic gas drag and planet-disk interactions, including $e$- and $i$-damping due to gravitation torque exerted by the disk. The treatments were identical to those presented in \cite{Lau2024b} and are summarized in Sect. \ref{sec:m:mig}.

\subsubsection{Numerical setup} \label{sec:m:num}
\begin{deluxetable*}{llllllllll}
	\tablewidth{0pt}
		\tablecaption{Summary of Parameters, Median Times of Key Event, and Median Final Mass of Kuiper Belt\label{tab:param}}
	\tablehead{
		\multicolumn{3}{c}{Parameter} & \multicolumn{6}{|c}{Result} \\
		\cline{1-3}  \cline{4-10}
		\colhead{$L_\mathrm{X}$} & \colhead{$M_\mathrm{disk}$} & \colhead{$\kappa$} & \colhead{Figure} & \colhead{1st gen.} & \colhead{2nd gen.} & \colhead{Cavity} & \colhead{Inner disk}& \colhead{Outer disk}& \colhead{Kuiper Belt}\\[-2ex]
		\colhead{} & \colhead{} & \colhead{} & \colhead{} & \colhead{of plts} & \colhead{of plts} & \colhead{Opening} & \colhead{depletion}& \colhead{depletion}& \colhead{mass}\\[-1ex]
		\colhead{$(\mathrm{erg\ s}^{-1})$} & \colhead{$(M_\odot)$} & \colhead{$(\mathrm{cm}^2\ \mathrm{g}^{-1})$} & \colhead{} & \colhead{(Myr)} & \colhead{(Myr)} & \colhead{(Myr)} & \colhead{(Myr)}& \colhead{(Myr)}& \colhead{($M_\oplus$)}
	}
	\startdata
	$10^{30}$ 	     & $0.025$ & 0.02 & Fig. \ref{fig:frames}, \ref{fig:plts_hist} \& \ref{fig:plts_peas}a & 0.13 & 0.53 & 0.49 &  0.93 & 1.87 & 2.23\\
	$5\times10^{29}$ & $0.025$ & 0.02 & Fig. \ref{fig:plts_peas}b & 0.14 & 0.70 & 0.82 & 1.42 & 3.14  & 0.183\\
	$2\times10^{29}$ & $0.025$ & 0.02 & Fig. \ref{fig:plts_peas}c & 0.15 & 0.91 & 2.55 & 2.69 & >5.00 & 0.0138\\
	$10^{30}$ 	     & $0.05$  & 0.02 & Fig. \ref{fig:plts_peas}d & 0.14 & 0.50 & 0.71 & 1.52 & 3.72  & 0.00789\\
	$10^{30}$ 	     & $0.025$ & 0.01 & Fig. \ref{fig:plts_peas}e & 0.13 & 0.48 & 0.44 & 0.92 & 1.88  & 2.01\\
	\enddata\tablecomments{The first set of parameters corresponds to the fiducial simulations. For all simulations, we considered a disk with $r_\mathrm{c}=50\ \mathrm{au}$, $\alpha=5\times10^{-4}$ around a solar-mass star. The initial pressure bump is described by eq. (\ref{eq:f_bump}) with $A=2$, $r_0=6$ au and $w=2H_\mathrm{g}(r=r_0)$. Five random simulations were conducted for each set of parameters. The final mass of Kuiper Belt is defined as the total mass of planetesimals ($\leq0.1M_\oplus$) with semimajor axis beyond that of the outermost giant planet ($>10M_\oplus$).}
\end{deluxetable*}

The fixed $N$-body timestep of $\tau=0.2$ years for SyMBAp was used and the particles were removed if the heliocentric distance is less than 2 au or greater than $576.8$ au. This range was set by trimming both ends of the radial grid of DustPy by 5 cells. In the fiducial case described later in this subsection, typically about 500 to 2,000 particles are removed at the inner boundary and about 2,000 to 3,300 particles are removed at the outer boundary throughout the simulation. The effect of implantation of planetesimals in the asteroid belt and planet formation in the terrestrial region will be studied in future works; therefore, the inner boundary is set to focus on the outer solar system in this work. The additional effects for the $N$-body particles were added to SyMBAp as
\begin{equation}
	\mathcal{P}^{\tau/2}\mathcal{M}^{\tau/2}\mathcal{N}^{\tau}\mathcal{M}^{\tau/2}\mathcal{P}^{\tau/2},
\end{equation}
where $\tau$ is the $N$-body timestep and each term is an operator described below. The operator $\mathcal{P}$ evolves the masses by pebble accretion, gas accretion and evaluates gap opening, $\mathcal{M}$ evolves the velocities by gas drag and planet migration, and $\mathcal{N}$ is the second-order symplectic integrator in SyMBAp. The operators $\mathcal{P}$ and $\mathcal{M}$ operate in the heliocentric coordinates and $\mathcal{N}$ operates in the democratic heliocentric coordinates, therefore, coordinate transformation was required at each step. The timestep management to couple SyMBAp and DustPy is detailed in Sect. 2.5 of \cite{Lau2022}.

We note a numerical difficulty when multiple giant planets were produced, which generally happened if the disk was relatively long-living in our test simulations. The presence of multiple steep slopes in the surface density greatly shortened the disk-integration timestep for DustPy. Since this work focuses on the final stage of the protoplanetary disk, the chosen values of $L_\mathrm{X}$ are in the high end of observation as mentioned in Sect. \ref{sec:m:PE}. The fiducial case has $L_\mathrm{X} = 10^{30}\ \mathrm{erg\ s}^{-1}$, $M_\mathrm{disk}=0.025\ M_\odot$ and $\kappa=0.02 \ \mathrm{cm}^2\ \mathrm{g}^{-1}$. We also tested $L_\mathrm{X} = \{2\times10^{29}, 5\times10^{29}\} \ \mathrm{erg\ s}^{-1}$, $M_\mathrm{disk}=0.05\ M_\odot$ and $\kappa=0.01 \ \mathrm{cm}^2\ \mathrm{g}^{-1}$. The whole set of parameters is summarized in the first 3 columns of Table \ref{tab:param}. Five simulations were conducted for each combination of parameters to evaluate the statistical effect. The simulation time was 5 Myr and each run required a wall-clock time of about 2 to 3 months. A total of 25 simulations were conducted and presented in the following section.

\section{Results} \label{sec:results}
\subsection{Fiducial simulations} \label{sec:fiducial}
\begin{figure*}
	\centering
	\includegraphics{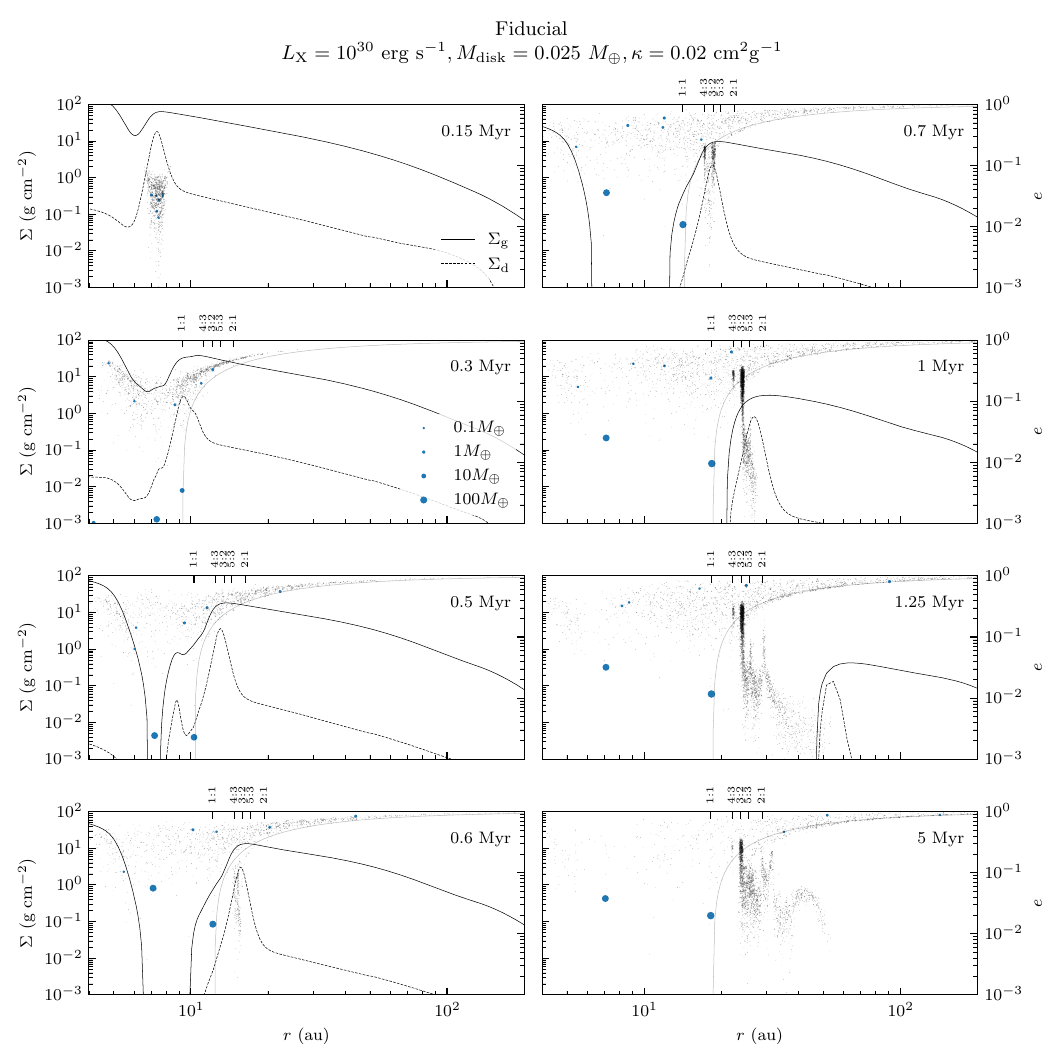}
	\caption{Eight key timestamps of one of the fiducial simulations. In each panel, the snapshot at the denoted time is shown. The solid and dashed lines, respectively, show the gas surface density $\Sigma_{\mathrm{g}}$ and dust surface density $\Sigma_{\mathrm{d}}$ with respect to the distance from the star $r$. The eccentricity $e$ and semimajor axis $r$ of the $N$-body particles are shown by the black dots, for mass $m\leq0.1M_\oplus$, and, otherwise, blue circles with the linear sizes proportional to $m^{1/3}$ and the provided scale. Upon the formation of massive planets, selected locations of mean motion commensurability with respect to the outermost one are also denoted in the top axis. And, the gray line in $r-e$ traces the orbits that just cross the outermost planet's apastron.
		\label{fig:frames}}
\end{figure*}
\begin{figure}
	\centering
	\includegraphics{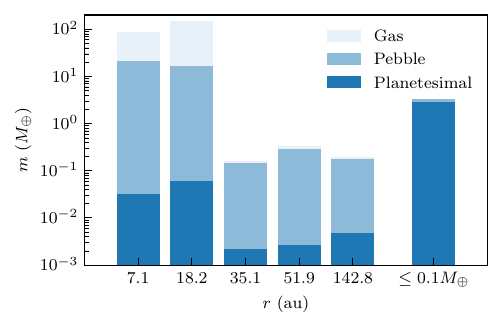}
	\caption{Origins of mass of the $N$-body particles at the end of the fiducial simulation. `Planetesimal' refers to the mass from the initial planetesimal formation and subsequent planetesimal accretions, `Pebble' refers to the mass accreted from the dust in the disk and `Gas' refers to the mass accreted from the gas in the disk. For bodies above $0.1M_\oplus$, the individual compositions are shown with the semimajor axis $r$ shown on the horizontal axis. For the remaining 3,283 minor bodies ($\leq0.1M_\oplus$), the sum of their compositions are shown instead, which is noted by `$\leq0.1M_\oplus$' on the horizontal axis.
		\label{fig:plts_compo}}
\end{figure}
\begin{figure}
	\centering
	\includegraphics{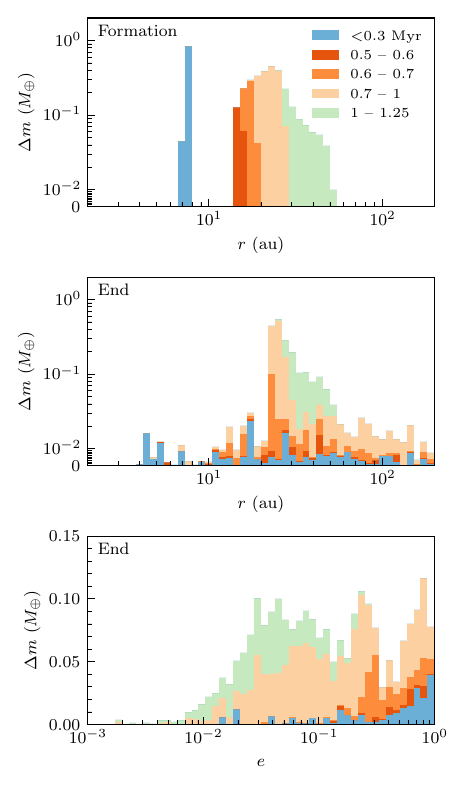}
	\caption{Distribution of planetesimal mass $m$ at formation in distance from the star $r$ (top), at the end of the fiducial simulation in Fig. \ref{fig:frames} (middle), and that in eccentricity $e$ at the end of the same simulation (bottom). The colors denote the formation time. The mass bins are log-uniform in the horizontal axis. Only the mass from planetesimal formation is considered while pebble accretion and planetesimal accretion are not shown here.
		\label{fig:plts_hist}}
\end{figure}
\begin{figure*}
	\centering
	\includegraphics{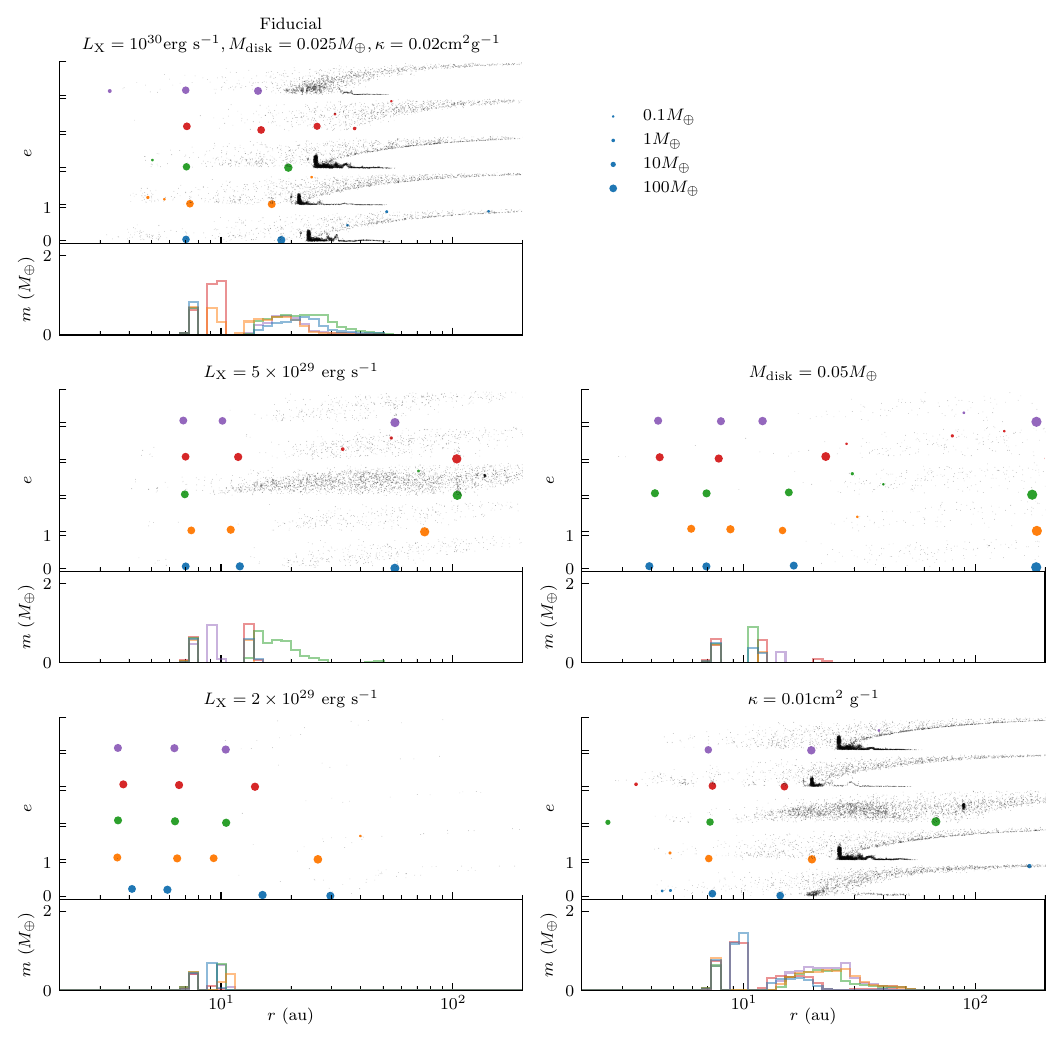}
	\caption{Results of all random simulations. Each sub-figure presents the results from the simulation with the denoted set of parameters: \textbf{a)} fiducial parameters; \textbf{b)} lower stellar X-ray luminosities $L_\mathrm{X}=5\times10^{29}\mathrm{\ erg\ s}^{-1}$; \textbf{c)} $L_\mathrm{X}=2\times10^{29}\mathrm{\ erg\ s}^{-1}$; \textbf{d)} doubled disk mass $M_\mathrm{disk}=0.05 M_\odot$; \textbf{e)} halved gas envelope opacity $\kappa=0.01 \ \mathrm{cm}^2\ \mathrm{g}^{-1}$. In each sub-figure, the top panel shows the semimajor axis $r$ and eccentricity $e$ of the particles for each random simulation, respectively. Similar to Fig. \ref{fig:frames}, the eccentricity $e$ and semimajor axis $r$ of the $N$-body particles are shown by the black dots, for mass $m\leq0.1M_\oplus$, and, otherwise, colored circles with the linear sizes proportional to $m^{1/3}$ and the provided scale. The bottom panel shows the radial distribution of the planetesimal mass at formation, with the colors corresponding to the random simulations in the top panel. The simulation denoted by the color blue in sub-figure a)  corresponds to the one presented in Fig. \ref{fig:frames} \& \ref{fig:plts_hist}. \label{fig:plts_peas}}
\end{figure*}

\subsubsection{Formation and evolution of minor planetary bodies} \label{sec:form}
Figure \ref{fig:frames} presents one of the fiducial simulations with the panels showing the 8 key timestamps noted inside each panel. The solid and dashed lines show the profiles of the gas surface density $\Sigma_\mathrm{g}$ and the dust surface density $\Sigma_\mathrm{d}$, respectively. The dots denote the eccentricity $e$ and the semimajor axis $r$ of the massive bodies up to $0.1M_\oplus$. Massive bodies above $0.1M_\oplus$ are shown by blue circles with the linear sizes proportional to $m^{1/3}$ and the provided scale.

At 0.15 Myr, planetesimals had formed at the imposed initial substructure, and rapid growth by pebble accretion had started, which is similar to the case previously presented in \cite{Lau2022,Lau2024b}. Through gravitational interactions, the smaller planetesimals were also being scattered from their initial locations and reached a significant eccentricity ($\gtrsim0.01$), while $e$-damping by the disk is insignificant in this mass regime of planetesimals as shown by eq. (\ref{eq:tau_e}). Pebble accretion became inefficient for these excited bodies due to the increased relative velocity with respect to the pebble.

At 0.3 Myr, two planetary cores had formed and started to perturb the disk. The locations of mean motion commensurability with respect to the outermost core are also denoted in the top axis, including the ratios of 1:1, 4:3, 3:2, 5:3, and 2:1. The gray line traces the orbits that would just cross the outermost planet's apastron. The smaller planetesimals experienced a much more vigorous scattering mainly to the outer disk as they have generally reached high eccentricities ($\gtrsim0.1$).

At 0.5 Myr, a cavity had already opened due to the rapid photoevaporation near the location of the inner giant planet, which is at about 7 au. Meanwhile, the outer giant planet was at about 10 au due to the gravitational interaction with the inner one. Scattering of the smaller planetesimals continued and reached both the inner and outer boundaries of the simulation domain. A ring of concentrated dust had formed external to the outer giant planet as the pressure maxima had been collecting dust drifting from the outer disk and also from the initial dust trap.

At 0.6 Myr, the cavity continued to expand and planetesimal formation resumed in the dust trap that moved along with the outer cavity wall. The inner gas giant had an increased eccentricity as it was no longer damped by the disk, while the gravitational interaction with the outer gas giant continued. As the location of the migration trap moved outwards with the expanding cavity, the outer gas giant migrated with the cavity wall. This resulted in the capture of the newly formed planetesimals in the resonance locations of the outer gas giant.

At 0.7 Myr, the outer gas giant continued to migrate with the retreating cavity wall while new planetesimals were forming. As the peak of the dust trap was closer to the location of 3:2 mean motion commensurability with respect to the outer gas giant, a larger population of planetesimals was captured there relative to the location of 4:3 mean motion commensurability.

At 1 Myr, the inner disk was depleted. The outer cavity wall was then under direct irradiation of the central star and its retreat became faster. Combined with a weakened migration due to the lower gas surface density, the outer gas giant had stopped outward migration and remained at about 18 au. This is mainly due to the torque exerted onto the disk by the planet has become stronger as the planet grew. Meanwhile, planetesimal formation continued at the outward-moving dust trap. The newly formed planetesimals were farther from the outer gas giant and experienced weaker dynamical stirring.

At 1.25 Myr, the outer cavity wall was at about 50 au and planetesimal formation had just terminated as the remaining dust mass was not able to create a dust concentration that could fulfill the formation criteria. The simulation resulted in a wide disk of planetesimals of about 30 au wide that traced the retreat of the cavity wall. In addition, as the later formed planetesimals were more distant from the giant planets, they were dynamically colder and not always captured into mean motion resonances, except the locations of 5:3 and 2:1 mean motion commensurability with respect to the outer gas giant.

The disk was eventually depleted at 1.6 Myr and the final panel shows the result of dynamical evolution up to 5 Myr. There were three distinct dynamical classes of small bodies. First, there was a class of small bodies that were scattered throughout the simulation domain and generally exhibited a high eccentricity of 0.1 and above. Second, there was another class that was captured in the resonance locations of the outer gas giant. Finally, there remained a class that the small bodies were dynamically cold with eccentricity generally less than 0.1.

\subsubsection{Origin of accreted mass} \label{sec:PA}
Figure \ref{fig:plts_compo} shows the origins of accreted mass individually for bodies above $0.1M_\oplus$ and the sum for all bodies  $\leq0.1M_\oplus$. The mass budget is separated into three parts: 1) `Planetesimal' refers to the mass from the formation of particles and the subsequent planetesimal accretions; 2) `Pebble' refers to the mass accreted from the dust in the disk, and; 3) `Gas' refers to the mass accreted from the gas in the disk. In case of a merger, the mass composition of the accreted bodies is added to the respective components. However, only 146 mergers occurred throughout the simulation, which contributed a negligible amount of gas and pebble mass to the planets. There is a distinction that pebble accretion contributed largely to the solid mass of planets while is a small fraction of the small bodies.

\subsubsection{Origin of minor planetary bodies} \label{sec:origin}
Figure \ref{fig:plts_hist} presents the distribution of the planetesimal mass at formation and at the end of the simulations with the formation time denoted by the colors. The top panel shows the distribution of semimajor axes when the planetesimals were first formed. It shows a clear trend that planetesimals formed at a later time were also formed at a larger distance from the star, with a temporal and spatial gap between those formed in the initial pressure bump and those formed at the cavity wall.

The middle and lower panels of Fig. \ref{fig:plts_hist} show the distribution of semimajor axis and eccentricity at the end of the simulation, respectively. Only the initial mass is considered while the mass gained from accretion is not included to focus on the relocation of particles. Throughout the simulation, only 135 mergers occurred among 5480 planetesimals that were formed, which mainly were collisions with the giant planets, and were not significant to this distribution. The planetesimals formed from the initial pressure bump, also formed before 0.3 Myr, were scattered throughout the simulation domain and generally had high eccentricities $\gtrsim0.3$.

For the planetesimals formed at the cavity wall, the earliest ones (0.5 to 0.6 Myr) also experienced significant scattering and reached high eccentricities. For those formed from 0.6 to 0.7 Myr, about half of the mass was captured in resonance and transported outward with the outer gas giant with elevated eccentricities of $\sim0.3$. As its outward migration stopped, a significant fraction of the planetesimals formed from 0.7 to 1 Myr remained in situ with relatively low eccentricities $\lesssim0.1$ while the rest were mostly scattered to the outer disk. Finally, the planetesimals formed from 1 to 1.25 Myr mostly remained in situ and dynamically cold with eccentricities $\lesssim0.1$.

\subsubsection{Stochasticity} \label{sec:rand}
Figure \ref{fig:plts_peas}a shows the end results of the five random simulations. The top panel shows the semimajor axis and eccentricity of the particles for each simulation, respectively. Similar to Fig. \ref{fig:frames}, the black dots correspond to the massive bodies up to $0.1M_\oplus$ and those above $0.1M_\oplus$ are shown by colored circles with the linear sizes proportional to $m^{1/3}$. The bottom panel shows the radial distribution of the planetesimal mass at formation and the colors correspond to the random simulations in the top panel. The simulation denoted by the color blue corresponds to the one presented in Fig. \ref{fig:frames} \& \ref{fig:plts_hist}.

Across the random simulations, three simulations (blue, orange, and green) showed similar end results that three classes of minor planetary bodies were formed. The simulation with the color red formed another generation of planets instead of the planetesimal belt, as only one giant planet was formed in the first generation. And, the simulation with the color purple had a less distinct class of resonant bodies as a planet was ejected outward and perturbed the outer planetary system. Despite the stochasticity, all random simulations produced a class of scattered planetesimals.

\subsection{Decreased X-ray luminosity of the star $L_\mathrm{X}$} \label{sec:Lx}
Figure \ref{fig:plts_peas}b \& c show the end results of the simulations with lower stellar X-ray luminosities $L_\mathrm{X}$ of $5\times10^{29}$ and $2\times10^{29}\,\mathrm{erg\ s}^{-1}$, respectively, in the same manner as in Fig. \ref{fig:plts_peas}a. In Table \ref{tab:param}, the five columns on the right present the median times of the key event for each set of parameters, respectively. The key events are the formation of the first and second generation of planetesimals, cavity opening, depletion of the inner disk, and depletion of the outer disk. The rightmost column presents the median of the final masses of the Kuiper Belt analogs, which are defined as the total mass of the planetesimals ($\leq0.1M_\oplus$) with semimajor axis beyond that of the outermost giant planet ($>10M_\oplus$) evaluated at the end of the simulations.

In the case of $L_\mathrm{X} = 5\times10^{29}\mathrm{\ erg\ s}^{-1}$ (Fig. \ref{fig:plts_peas}b), planetesimals formed at the planetary gap of the first generation giant planets at 0.70 Myr before the cavity opened at 0.82 Myr. Since the formation of this second generation of planetesimals was still early in terms of the disk's lifetime, a giant planet was formed there and experienced significant outward migration as the cavity opened and widened. However, there was not enough dust left to further form planetesimals. Only the planetesimals formed with and scattered by the giant planets remained at the end of the simulations. The final mass of the Kuiper Belt was about $0.183M_\oplus$, which is an order of magnitude lower than that of the fiducial simulations.

In the case of $L_\mathrm{X} = 2\times10^{29}\mathrm{\ erg\ s}^{-1}$ (Fig. \ref{fig:plts_peas}c), the formation of giant planets was even faster as planetesimals grew to planetary cores in a shorter time. As the dust was depleted quickly by pebble accretion of the cores, significantly fewer planetesimals were formed, resulting in significantly fewer small bodies remaining at the end of the simulations. Since the gas surface density was lower when the cavity opened at 2.55 Myr, outward migration of the outermost planet was less significant. A subsequent dynamical instability occurred among the compact chain of planets. The final mass of the Kuiper Belt was even lower, about $0.0138M_\oplus$.

\subsection{Doubled disk mass $M_\mathrm{disk}$} \label{sec:mdisk}
Figure \ref{fig:plts_peas}d shows the end results of the simulations with a doubled disk mass $M_\mathrm{disk}=0.05 M_\odot$. The formation of gas giants in the initial pressure bump occurred earlier compared to the fiducial simulations, as the growth by pebble accretion is rapid. Similar to the case of decreased $L_\mathrm{X}$, a second generation of giant planets was formed at 0.50 Myr from the gap opened by the first generation of giant planets. Also, fewer planetesimals were formed due to efficient pebble accretion. The end result is similar to that of $L_\mathrm{X} = 2\times10^{29}\mathrm{\ erg\ s}^{-1}$ (Fig. \ref{fig:plts_peas}c) with generally one more giant planet in the second generation and the final mass of the Kuiper Belt approximately halved. This emphasizes the interplay of planetesimal formation and pebble accretion in consuming dust mass.

\subsection{Halved envelope opacity $\kappa$} \label{sec:kappa}
Figure \ref{fig:plts_peas}e shows the end results of the simulations with a halved gas envelope opacity $\kappa=0.01 \ \mathrm{cm}^2\ \mathrm{g}^{-1}$. The end results are very similar to those of the fiducial simulations (Fig.~\ref{fig:plts_peas}a). This is due to the short thermal contraction phase of gas accretion ($\sim10^4$ to $10^5$ yr) in both cases and the change in the value of $\kappa$ is insignificant relative to the evolution of the disk.

\section{Discussions} \label{sec:dis}
\subsection{Formation of dynamical groups of minor planetary bodies}
\begin{figure}
	\centering
	\includegraphics{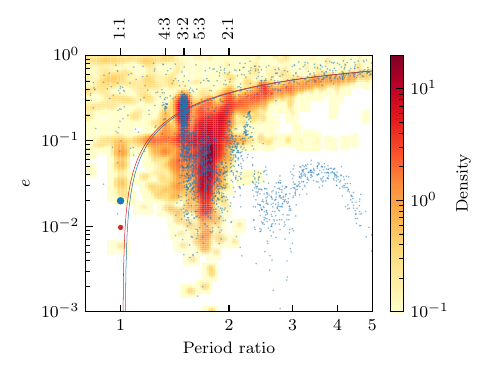}
	\caption{Comparison of the fiducial simulation result with the observed Kuiper Belt objects retrieved from the Minor Planet Center on 2025 May 17. The two sets of data are overlayed in a manner that the horizontal axis shows the period ratio with respect to the outermost giant planet of 145.8 $M_\oplus$ (blue circle) in our simulation and Neptune (red circle), respectively. The color map presents the number density of Kuiper Belt objects, and the blue dots present the result of our fiducial simulation as shown in the last frame of Fig. \ref{fig:frames}. The linear sizes of the circles are proportional to $m^{1/3}$. The blue and red lines trace the orbits that just cross outermost giant planet's apastron and Neptune's aphelion, respectively. Our result reproduced the major dynamical groups: the scattered group, the resonant group, and the dynamically cold group. 		
		\label{fig:tno}}
\end{figure}
The fiducial simulations (Sect. \ref{sec:fiducial}) demonstrated a scenario of the formation of three distinct dynamical groups of minor planetary bodies. The scattered group, which has the highest eccentricity ($e\gtrsim0.4$), consists of planetesimals that were formed with the giant planets and a fraction of those formed at the cavity wall while the outer planet was still migrating outward. Most of the mass in this class was scattered external to the giant planets, with only less than 0.05 $M_\oplus$ ending up inside the orbit of the inner gas giant.

The resonant group, which was in mean motion resonance with the outer gas giant, had formed continuously at the cavity wall and was captured by the outward migrating planet. Although these bodies remained near the dust ring, they were excited to significant eccentricities ($e\gtrsim0.1$) that greatly increased the pebble accretion onset mass. As discussed in \cite{Lau2022,Lau2024}, the rate of pebble accretion is sensitive to the relative velocity and the dynamical stirring of small bodies can effectively prevent them from accreting pebbles. In terms of the origin, this class of bodies experienced significant outward transportation, especially those formed earlier and closer to the star, as they were captured in resonance with the outermost giant planet.

Finally, the dynamically cold group was formed at the cavity wall after the giant planet had stopped migrating outward. Although they were dynamically cold, the cavity wall and the dust ring moved beyond them quickly due to photoevaporation, such that growth by pebble accretion was not significant. Planet migration is also insignificant in this mass regime. Therefore, their locations and masses generally remained the same as they were formed.

In the context of the solar system, Fig. \ref{fig:tno} compares the orbits of known Kuiper Belt objects and the simulation result. Our fiducial simulation (Fig. \ref{fig:frames}) closely reproduced the three major dynamical groups when the radial distances are dynamically scaled with respect to the outermost planet or Neptune. Nonetheless, the simulation result showed additional objects formed farther than the location of 2:1 mean motion commensurability, while similar objects are not observed in the Kuiper Belt. Also, the current estimate of the Kuiper Belt's mass is at $\sim 0.2M_\oplus$ \citep{Pitjeva2018} while the simulation results in about $2.23 M_\oplus$ of small bodies (Table \ref{tab:param}). Long-term evolution ($\sim$ Gyr) studies and modeling the formation of the giant planets in the solar system are required to model the loss rate and to further confirm this formation scenario.

Our results suggest that the dynamical properties of the orbits of minor planetary bodies have a strong correlation with their formation time and location. In particular, the growth by pebble accretion is sensitive to dynamics while planetesimals are formed in dust-rich regions, which also favors pebble accretion. The presence of minor planetary bodies suggests a specific formation scenario. This can provide an important constraint on the formation and dynamical history of a planetary system. This includes both the Kuiper Belt's architecture as well as the observations of exo-Kuiper belts, where the thickness of debris disk and the sharpness of edges provide constraints on their dynamical properties \citep[e.g.][]{Daley2019,Marino2021,ImazBlanco2023,Matra2025}.

On the other hand, our parameter study (Sect. \ref{sec:Lx} to \ref{sec:kappa}) showed that these groups of minor bodies might not be universally present in a planetary system. This is further discussed below.

\subsection{Competition between planetesimal formation and pebble accretion}\label{dis:SI_PA}
The fiducial simulations generally resulted in an external planetesimal belt of about 2$M_\oplus$, which did not grow significantly by pebble accretion but the planets did so (Fig. \ref{fig:plts_compo}). In contrast, the results of slower photoevaporation (Sect. \ref{sec:Lx}) show that when the planetesimals formed before cavity opening, a second generation of giant planets could form instead due to efficient pebble accretion. In the case of $L_\mathrm{X} = 5\times10^{29}\mathrm{\ erg\ s}^{-1}$ (Fig. \ref{fig:plts_peas}b), the cavity opened soon after the giant planet had formed from the new pressure bump. This terminated the formation of more giant planets and the planetesimals present there were scattered. As a significant fraction of dust had been accreted through pebble accretion, planetesimals were not formed at the cavity wall. Therefore, the resonant group and dynamical group of minor bodies present in the fiducial simulations were absent in this case. Combined with significant outward migration of the giant planets, this resulted in a much less massive Kuiper Belt ($0.183M_\oplus$). In the case of $L_\mathrm{X} = 2\times10^{29}\mathrm{\ erg\ s}^{-1}$ (Fig. \ref{fig:plts_peas}c), two planetary cores could form rapidly in the new pressure bump before the cavity opened. Therefore, pebble accretion was even more efficient than planetesimal formation in consuming dust mass in this case. Planetesimal formation was suppressed further with $\sim 10^{-2} M_\oplus$ of planetesimals left beyond the giant planets at the end of the simulation. Similarly, with a higher disk mass (Sect. \ref{sec:mdisk}), rapid formation of planetary cores also suppressed planetesimal formation. This case resulted in the least planetesimal mass (0.00789 $M_\oplus$) external to the planet, despite a more massive disk, also due to the significant outward migration of the gas giants.

Although different disk sizes were not explored in this work, we expect that for a larger $R_\mathrm{c}$, dust in the outer disk might not have grown and drifted significantly when the cavity has opened. This would help preserving dust mass and favor planetesimal formation during disk dissipation.

We note that this result of the direct competition between planetesimal formation and pebble accretion should be sensitive to planetesimal formation efficiency. However, we have only considered the streaming instability and the chosen value of $\zeta = 10^{-3}$ is a conservative choice compared to other models \citep[e.g.][]{Schoonenberg2018,Miller2021}. This disfavors planetesimal formation, while the value of $\zeta$ is not well-constrained from the current local simulations of streaming instability. It requires further investigations, including a broader parameter study with our model for both smaller and larger values of $\zeta$. For instance, to reproduce the initial conditions for the Nice model, pebble accretion would have to be much less efficient than planetesimal formation in the outer solar system to produce a massive primordial Kuiper Belt of more than $10M_\oplus$ without forming more planets, which would require extreme parameters in our model. An alternative way to drive the outward migration of Neptune is proposed below.

\subsection{Outward migration during disk dissipation}
The results generally demonstrated outward migration of the outermost giant planet with the outward-moving cavity wall. In the cases of rapid disk dispersal (Fig. \ref{fig:plts_peas}a \& e), the cavity wall retreated faster than planet migration, especially when the inner disk was depleted, so the planet eventually left the disk. Meanwhile, in the cases of slow disk dispersal (Fig. \ref{fig:plts_peas}c), the migration strength was weaker when the cavity opened due to the low gas surface density and outward migration was less significant. In the intermediate case (Fig. \ref{fig:plts_peas}b \& d), outward migration was the most prominent as the outer planet followed the disk cavity closely. This provided a mechanism to transport planets far away from the star, while the number is likely limited to one per system, as the zero-torque region is a specific location near the cavity wall. Nonetheless, external photoevaporation was not included in our model, which will likely limit the extent of the outer disk as well as that of outward migration.

We suggest that, instead of migrating through a massive planetesimal disk, the outward migration of Neptune could have been a combined effect of a less massive primordial Kuiper Belt and a retreating cavity wall of the solar nebula. This scenario prevents the problem that a massive planetesimal belt triggering the formation of another planet in the outer solar system as discussed in Sect. \ref{dis:SI_PA} above.

\subsection{Gas envelope opacity}
The results of halved envelope opacity (Fig. \ref{fig:plts_peas}e) are qualitatively similar to those of the fiducial simulations (Fig. \ref{fig:plts_peas}a). While the difference in the duration of the thermal contraction phase was insignificant compared to the overall timescale of disk evolution, this suggests that a wider range of $\kappa$ should be tested, given its uncertainty. In the limit of a slow gas accretion rate, dust might not be retained until cavity opening to produce planetesimals. Meanwhile, in the limit of fast gas accretion, multiple generations of gas giants might form within the disk lifetime and pebble accretion would dominate over planetesimal formation. These and the interplay with other parameters require further investigations to verify.

\subsection{Limitations and outlooks}
The adopted parameter space was limited due to the high computational cost and we could only model short-lived disks that disperse around 1 Myr. In the context of the solar system, meteoritic records show that planetesimal formation likely had lasted for at least a few Myr, which sets the lower limit of the solar nebula's lifetime. The location of the initial pressure bump and the disk size were not explored either, but they are likely critical to the result. Similar to the difficulty discussed in \cite{Lau2024b}, modeling the disk is the bottleneck and we expect that it will be overcome with the implementation of TriPod \citep{Pfeil2024}, an efficient dust evolution model. Combining sequential giant planet formation and this model of the planetesimal formation during disk dispersal, this provides a promising pathway to model the formation of the outer solar system. Our model can also seamlessly be extended to study the long-term evolution of the planetary system after disk dispersal, while this is outside of the scope of this particular work. In the meantime, this improvement shall allow for a wider parameter study to be compared against the observed exoplanetary systems. Finally, we noted a degree of stochasticity in our results, which shall be better understood with a larger number of random simulations. In particular, combining with the sequential planet formation proposed in \cite{Lau2024b}, this model shows a promising pathway for a formation scenario of the outer solar system.

\section{Conclusions} \label{sec:concl}
This work demonstrates a scenario of the formation of multiple dynamical classes of minor planetary bodies during disk dissipation. We further extended the model in \cite{Lau2022} \& \cite{Lau2024b} by including internal photoevaporation to disperse the disk. In the fiducial case, three dynamical classes were formed (Fig. \ref{fig:frames} \& \ref{fig:plts_hist}a), which resemble the three corresponding classes of Kuiper Belt objects (Fig. \ref{fig:tno}). 

The dynamically hottest group ($e\gtrsim0.3$) corresponds to the planetesimals that were the siblings of the giant planets, but their growth was prevented due to dynamical stirring of the more massive cores that formed and grew earlier. Then, the resonant group corresponds to the planetesimals that were captured in resonance with the outermost giant planet. As a cavity opened and was expanding, a new dust trap and a migration trap were formed, both following the retreat of the cavity wall. The outward-migrating planet captured the planetesimals formed in the dust trap and transported them outward. This group had an elevated eccentricity of $\sim0.3$ that effectively prevented their growth by pebble accretion. When the inner disk had depleted, the outer disk was under direct irradiation of the star, which intensified the retreat of the cavity wall. Combined with the lower gas surface density at this stage, the planet could no longer catch up with the cavity wall and the outward migration stopped. A new class of planetesimals was formed after this point as they were not captured in resonance, which remained in-situ and dynamically cold ($e\lesssim0.1$) due to the distance from the giant planets. Due to the rapid retreat of the cavity wall, this group could not grow significantly by pebble accretion either. Given that planetesimal formation by the streaming instability requires concentrated dust, which is also a favorable condition for efficient pebble accretion, this scenario provided a pathway to form minor planetary bodies without significant growth.

We also showed that the formation of small bodies or all of the above three groups is not universal during disk dissipation. Our parameter study showed a competition for dust between planetesimal formation and pebble accretion by early-formed planetary cores. As discussed in Sect. \ref{dis:SI_PA}, forming a Kuiper Belt during disk dissipation requires enough remaining dust mass that was not accreted to form planetary cores. In the case of abundant dust mass or early core formation (Fig. \ref{fig:plts_peas}b to c), another generation of planets could form instead of planetesimals alone due to efficient pebble accretion. In the marginal case (Fig. \ref{fig:plts_peas}c), a scattered group of planetesimals could still remain after disk dispersal.

\begin{acknowledgments}
 T.C.H.L., T.B. and S.M.S. acknowledge funding from the European Union under the European Union's Horizon Europe Research and Innovation Programme 101124282 (EARLYBIRD) and funding by the Deutsche Forschungsgemeinschaft (DFG, German Research Foundation) under grant 325594231, and Germany's Excellence Strategy - EXC-2094 - 390783311. J.D. was funded by the European Union under the European Union’s Horizon Europe Research \& Innovation Programme 101040037 (PLANETOIDS). Views and opinions expressed are, however those of the author(s) only and do not necessarily reflect those of the European Union or the European Research Council. Neither the European Union nor the granting authority can be held responsible for them. This research has made use of data and/or services provided by the International Astronomical Union's Minor Planet Center.\\
\end{acknowledgments}

\bibliography{aas_PE}

@article{Adachi1976,
  title = {The Gas Drag Effect on the Elliptic Motion of a Solid Body in the Primordial Solar Nebula},
  author = {Adachi, Isao and Hayashi, Chushiro and Nakazawa, Kiyoshi},
  year = 1976,
  month = dec,
  journal = {PThPh},
  volume = {56},
  number = {6},
  eprint = {https://academic.oup.com/ptp/article-pdf/56/6/1756/5174571/56-6-1756.pdf},
  pages = {1756--1771},
  issn = {0033-068X},
  doi = {10.1143/PTP.56.1756}
}

@article{Armitage2002,
  title = {The Brown Dwarf Desert as a Consequence of Orbital Migration},
  author = {Armitage, Philip J. and Bonnell, Ian A.},
  year = 2002,
  month = feb,
  journal = {MNRAS},
  volume = {330},
  number = {1},
  pages = {L11-L14},
  issn = {0035-8711},
  doi = {10.1046/j.1365-8711.2002.05213.x},
  urldate = {2024-05-16}
}

@article{Bitsch2015,
  title = {The Growth of Planets by Pebble Accretion in Evolving Protoplanetary Discs{$\star$}},
  author = {Bitsch, Bertram and Lambrechts, Michiel and Johansen, Anders},
  year = 2015,
  month = oct,
  journal = {A\&A},
  volume = {582},
  __markedentry = {[tommy:6]},
  refid = {10.105100046361201526463},
  file = {/home/tommy/snap/zotero-snap/common/Zotero/storage/5PMNZ325/Bitsch et al. - 2015 - The growth of planets by pebble accretion in evolving protoplanetary discs⋆.pdf}
}

@article{Bitsch2019,
  title = {Formation of Planetary Systems by Pebble Accretion and Migration: {{Growth}} of Gas Giants},
  author = {Bitsch, Bertram and Izidoro, Andre and Johansen, Anders and Raymond, Sean N. and Morbidelli, Alessandro and Lambrechts, Michiel and Jacobson, Seth A.},
  year = 2019,
  journal = {A\&A},
  volume = {623},
  pages = {A88},
  doi = {10.1051/0004-6361/201834489}
}

@article{Bodenheimer1986,
  title = {Calculations of the Accretion and Evolution of Giant Planets: {{The}} Effects of Solid Cores},
  shorttitle = {Calculations of the Accretion and Evolution of Giant Planets},
  author = {Bodenheimer, Peter and Pollack, James B.},
  year = 1986,
  month = sep,
  journal = {Icarus},
  volume = {67},
  number = {3},
  pages = {391--408},
  issn = {0019-1035},
  doi = {10.1016/0019-1035(86)90122-3},
  urldate = {2024-08-19}
}

@article{Brouwers2021,
  title = {How Planets Grow by Pebble Accretion},
  author = {Brouwers, M. G. and Ormel, C. W. and Bonsor, A. and Vazan, A.},
  year = 2021,
  journal = {A\&A},
  volume = {653},
  doi = {10.1051/0004-6361/202140476},
  refid = {10.105100046361202140476}
}

@article{Carrera2015,
  title = {How to Form Planetesimals from Mm-Sized Chondrules and Chondrule Aggregates},
  author = {Carrera, Daniel and Johansen, Anders and Davies, Melvyn B.},
  year = 2015,
  month = jul,
  journal = {A\&A},
  volume = {579},
  pages = {A43},
  publisher = {EDP Sciences},
  issn = {0004-6361, 1432-0746},
  doi = {10.1051/0004-6361/201425120},
  urldate = {2025-03-21},
  copyright = {\copyright{} ESO, 2015},
  langid = {english}
}

@article{Carrera2017,
  title = {Planetesimal Formation by the Streaming Instability in a Photoevaporating Disk},
  author = {Carrera, Daniel and Gorti, Uma and Johansen, Anders and Davies, Melvyn B.},
  year = 2017,
  month = apr,
  journal = {ApJ},
  volume = {839},
  number = {1},
  pages = {16},
  publisher = {The American Astronomical Society},
  issn = {0004-637X},
  doi = {10.3847/1538-4357/aa6932}
}

@article{Chambers2021,
  title = {Rapid {{Formation}} of {{Jupiter}} and {{Wide-orbit Exoplanets}} in {{Disks}} with {{Pressure Bumps}}},
  author = {Chambers, John},
  year = 2021,
  month = jun,
  journal = {ApJ},
  volume = {914},
  number = {2},
  pages = {102},
  publisher = {The American Astronomical Society},
  issn = {0004-637X},
  doi = {10.3847/1538-4357/abfaa4},
  urldate = {2024-12-19},
  langid = {english}
}

@article{Chang2023,
  title = {On the {{Origin}} of {{Dust Structures}} in {{Protoplanetary Disks}}: {{Constraints}} from the {{Rossby Wave Instability}}},
  shorttitle = {On the {{Origin}} of {{Dust Structures}} in {{Protoplanetary Disks}}},
  author = {Chang, Eonho and Youdin, Andrew N. and Krapp, Leonardo},
  year = 2023,
  month = mar,
  journal = {ApJL},
  volume = {946},
  number = {1},
  pages = {L1},
  publisher = {The American Astronomical Society},
  issn = {2041-8205},
  doi = {10.3847/2041-8213/acc17b},
  urldate = {2025-07-15},
  langid = {english}
}

@article{Clarke1988,
  title = {The Diffusion of Contaminant through an Accretion Disc},
  author = {Clarke, C. J. and Pringle, J. E.},
  year = 1988,
  month = nov,
  journal = {MNRAS},
  volume = {235},
  number = {2},
  eprint = {https://academic.oup.com/mnras/article-pdf/235/2/365/3048056/mnras235-0365.pdf},
  pages = {365--373},
  issn = {0035-8711},
  doi = {10.1093/mnras/235.2.365}
}

@article{Clement2017,
  title = {Prevalence of Chaos in Planetary Systems Formed through Embryo Accretion},
  author = {Clement, Matthew S. and Kaib, Nathan A.},
  year = 2017,
  month = may,
  journal = {Icarus},
  volume = {288},
  pages = {88--98},
  issn = {0019-1035},
  doi = {10.1016/j.icarus.2017.01.021},
  urldate = {2024-12-19}
}

@article{Daley2019,
  title = {The {{Mass}} of {{Stirring Bodies}} in the {{AU Mic Debris Disk Inferred}} from {{Resolved Vertical Structure}}},
  author = {Daley, Cail and Hughes, A. Meredith and Carter, Evan S. and Flaherty, Kevin and Lambros, Zachary and Pan, Margaret and Schlichting, Hilke and Chiang, Eugene and Wyatt, Mark and Wilner, David and Andrews, Sean and Carpenter, John},
  year = 2019,
  month = apr,
  journal = {ApJ},
  volume = {875},
  number = {2},
  pages = {87},
  publisher = {The American Astronomical Society},
  issn = {0004-637X},
  doi = {10.3847/1538-4357/ab1074},
  urldate = {2025-07-15},
  langid = {english}
}

@article{DAngelo2010,
  title = {{{THREE-DIMENSIONAL DISK}}--{{PLANET TORQUES IN A LOCALLY ISOTHERMAL DISK}}},
  author = {D'Angelo, Gennaro and Lubow, Stephen H.},
  year = 2010,
  month = nov,
  journal = {ApJ},
  volume = {724},
  number = {1},
  pages = {730},
  publisher = {The American Astronomical Society},
  issn = {0004-637X},
  doi = {10.1088/0004-637X/724/1/730},
  urldate = {2025-03-19},
  langid = {english}
}

@article{Deienno2018,
  title = {Excitation of a {{Primordial Cold Asteroid Belt}} as an {{Outcome}} of {{Planetary Instability}}},
  author = {Deienno, Rogerio and Izidoro, Andr{\'e} and Morbidelli, Alessandro and Gomes, Rodney S. and Nesvorn{\'y}, David and Raymond, Sean N.},
  year = 2018,
  month = aug,
  journal = {ApJ},
  volume = {864},
  number = {1},
  pages = {50},
  publisher = {The American Astronomical Society},
  issn = {0004-637X},
  doi = {10.3847/1538-4357/aad55d},
  urldate = {2024-12-19},
  langid = {english}
}

@article{Drazkowska2016,
  title = {Close-in Planetesimal Formation by Pile-up of Drifting Pebbles},
  author = {Dr{\k a}{\.z}kowska, J. and Alibert, Y. and Moore, B.},
  year = 2016,
  journal = {A\&A},
  volume = {594},
  doi = {10.1051/0004-6361/201628983},
  refid = {10.105100046361201628983}
}

@article{Dubrulle1995,
  title = {The Dust Subdisk in the Protoplanetary Nebula},
  author = {Dubrulle, B. and Morfill, G. and Sterzik, M.},
  year = 1995,
  journal = {Icar},
  volume = {114},
  number = {2},
  pages = {237--246},
  issn = {0019-1035},
  doi = {10.1006/icar.1995.1058}
}

@article{Duffell2015,
  title = {{{HALTING MIGRATION}}: {{NUMERICAL CALCULATIONS OF COROTATION TORQUES IN THE WEAKLY NONLINEAR REGIME}}},
  shorttitle = {{{HALTING MIGRATION}}},
  author = {Duffell, Paul C.},
  year = 2015,
  month = jun,
  journal = {ApJ},
  volume = {806},
  number = {2},
  pages = {182},
  publisher = {The American Astronomical Society},
  issn = {0004-637X},
  doi = {10.1088/0004-637X/806/2/182},
  urldate = {2025-09-23},
  langid = {english}
}

@article{Duffell2020,
  title = {An {{Empirically Derived Formula}} for the {{Shape}} of {{Planet-induced Gaps}} in {{Protoplanetary Disks}}},
  author = {Duffell, Paul C.},
  year = 2020,
  month = jan,
  journal = {ApJ},
  volume = {889},
  number = {1},
  pages = {16},
  publisher = {The American Astronomical Society},
  issn = {0004-637X},
  doi = {10.3847/1538-4357/ab5b0f},
  urldate = {2024-12-19},
  langid = {english}
}

@article{Dullemond2018,
  title = {Dust-Driven Viscous Ring-Instability in Protoplanetary Disks},
  author = {Dullemond, C. P. and Penzlin, A. B. T.},
  year = 2018,
  month = jan,
  journal = {A\&A},
  volume = {609},
  pages = {A50},
  publisher = {EDP Sciences},
  issn = {0004-6361, 1432-0746},
  doi = {10.1051/0004-6361/201731878},
  urldate = {2024-04-10},
  copyright = {\copyright{} ESO, 2018},
  langid = {english}
}

@article{Dullemond2018a,
  title = {The Disk Substructures at High Angular Resolution Project ({{DSHARP}}). {{VI}}. {{Dust}} Trapping in Thin-Ringed Protoplanetary Disks},
  author = {Dullemond, Cornelis P. and Birnstiel, Tilman and Huang, Jane and Kurtovic, Nicol{\'a}s T. and Andrews, Sean M. and Guzm{\'a}n, Viviana V. and P{\'e}rez, Laura M. and Isella, Andrea and Zhu, Zhaohuan and Benisty, Myriam and Wilner, David J. and Bai, Xue-Ning and Carpenter, John M. and Zhang, Shangjia and Ricci, Luca},
  year = 2018,
  month = dec,
  journal = {ApJ},
  volume = {869},
  number = {2},
  pages = {L46},
  publisher = {American Astronomical Society},
  doi = {10.3847/2041-8213/aaf742}
}

@article{Duncan1998,
  title = {A Multiple Time Step Symplectic Algorithm for Integrating Close Encounters},
  author = {Duncan, Martin J. and Levison, Harold F. and Lee, Man Hoi},
  year = 1998,
  month = oct,
  journal = {AJ},
  volume = {116},
  number = {4},
  pages = {2067--2077},
  publisher = {IOP Publishing},
  doi = {10.1086/300541}
}

@article{Ercolano2017,
  title = {X-Ray Photoevaporation's Limited Success in the Formation of Planetesimals by the Streaming Instability},
  author = {Ercolano, Barbara and Jennings, Jeff and Rosotti, Giovanni and Birnstiel, Tilman},
  year = 2017,
  month = dec,
  journal = {MNRAS},
  volume = {472},
  number = {4},
  pages = {4117--4125},
  issn = {0035-8711},
  doi = {10.1093/mnras/stx2294},
  urldate = {2025-05-17}
}

@article{Flock2015,
  title = {Gaps, Rings, and Non-Axisymmetric Structures in Protoplanetary Disks - {{From}} Simulations to {{ALMA}} Observations},
  author = {Flock, M. and Ruge, J. P. and Dzyurkevich, N. and Henning, Th and Klahr, H. and Wolf, S.},
  year = 2015,
  month = feb,
  journal = {A\&A},
  volume = {574},
  pages = {A68},
  publisher = {EDP Sciences},
  issn = {0004-6361, 1432-0746},
  doi = {10.1051/0004-6361/201424693},
  urldate = {2024-04-10},
  copyright = {\copyright{} ESO, 2015},
  langid = {english}
}

@article{Franz2020,
  title = {Dust Entrainment in Photoevaporative Winds: {{The}} Impact of {{X-rays}}},
  shorttitle = {Dust Entrainment in Photoevaporative Winds},
  author = {Franz, R. and Picogna, G. and Ercolano, B. and Birnstiel, T.},
  year = 2020,
  month = mar,
  journal = {A\&A},
  volume = {635},
  pages = {A53},
  publisher = {EDP Sciences},
  issn = {0004-6361, 1432-0746},
  doi = {10.1051/0004-6361/201936615},
  urldate = {2025-03-21},
  copyright = {\copyright{} ESO 2020},
  langid = {english}
}

@article{Garate2021,
  title = {Large Gaps and High Accretion Rates in Photoevaporative Transition Disks with a Dead Zone},
  author = {G{\'a}rate, Mat{\'i}as and Delage, Timmy N. and Stadler, Jochen and Pinilla, Paola and Birnstiel, Til and Stammler, Sebastian Markus and Picogna, Giovanni and Ercolano, Barbara and Franz, Raphael and Lenz, Christian},
  year = 2021,
  month = nov,
  journal = {A\&A},
  volume = {655},
  pages = {A18},
  publisher = {EDP Sciences},
  issn = {0004-6361, 1432-0746},
  doi = {10.1051/0004-6361/202141444},
  urldate = {2024-12-19},
  copyright = {\copyright{} M. G\'arate et al. 2021},
  langid = {english}
}

@article{Gerbig2023,
  title = {Planetesimal {{Initial Mass Functions Following Diffusion-regulated Gravitational Collapse}}},
  author = {Gerbig, Konstantin and Li, Rixin},
  year = 2023,
  month = jun,
  journal = {ApJ},
  volume = {949},
  number = {2},
  pages = {81},
  publisher = {The American Astronomical Society},
  issn = {0004-637X},
  doi = {10.3847/1538-4357/acca1a},
  urldate = {2024-04-05},
  langid = {english}
}

@article{Gorti2015,
  title = {{{THE IMPACT OF DUST EVOLUTION AND PHOTOEVAPORATION ON DISK DISPERSAL}}},
  author = {Gorti, U. and Hollenbach, D. and Dullemond, C. P.},
  year = 2015,
  month = apr,
  journal = {ApJ},
  volume = {804},
  number = {1},
  pages = {29},
  publisher = {The American Astronomical Society},
  issn = {0004-637X},
  doi = {10.1088/0004-637X/804/1/29},
  urldate = {2025-07-15},
  langid = {english}
}

@incollection{Guilbert-Lepoutre2024,
  title = {Comet {{Nucleus Interiors}}},
  booktitle = {Comets {{III}}},
  author = {{Guilbert-Lepoutre}, Aur{\'e}lie and Davidsson, Bj{\"o}rn J. R. and Scheeres, Daniel J. and Ciarletti, Val{\'e}rie},
  year = 2024,
  month = jan,
  pages = {249--288},
  urldate = {2025-07-18}
}

@article{Gupta2023,
  title = {Reflections on Nebulae around Young Stars - {{A}} Systematic Search for Late-Stage Infall of Material onto {{Class II}} Disks},
  author = {Gupta, A. and Miotello, A. and Manara, C. F. and Williams, J. P. and Facchini, S. and Beccari, G. and Birnstiel, T. and Ginski, C. and Hacar, A. and K{\"u}ffmeier, M. and Testi, L. and Tychoniec, L. and Yen, H.-W.},
  year = 2023,
  month = feb,
  journal = {A\&A},
  volume = {670},
  pages = {L8},
  publisher = {EDP Sciences},
  issn = {0004-6361, 1432-0746},
  doi = {10.1051/0004-6361/202245254},
  urldate = {2024-04-10},
  copyright = {\copyright{} The Authors 2023},
  langid = {english}
}

@article{Ida2004,
  title = {Toward a Deterministic Model of Planetary Formation. {{I}}. {{A}} Desert in the Mass and Semimajor Axis Distributions of Extrasolar Planets},
  author = {Ida, S. and Lin, D. N. C.},
  year = 2004,
  month = mar,
  journal = {ApJ},
  volume = {604},
  number = {1},
  eprint = {astro-ph/0312144},
  pages = {388--413},
  doi = {10.1086/381724},
  archiveprefix = {arXiv}
}

@article{Ida2020,
  title = {A New and Simple Prescription for Planet Orbital Migration and Eccentricity Damping by Planet--Disc Interactions Based on Dynamical Friction},
  author = {Ida, Shigeru and Muto, Takayuki and Matsumura, Soko and Brasser, Ramon},
  year = 2020,
  month = may,
  journal = {MNRAS},
  volume = {494},
  number = {4},
  eprint = {https://academic.oup.com/mnras/article-pdf/494/4/5666/33205797/staa1073.pdf},
  pages = {5666--5674},
  issn = {0035-8711},
  doi = {10.1093/mnras/staa1073}
}

@article{Ikoma2000,
  title = {Formation of Giant Planets: {{Dependences}} on Core Accretion Rate and Grain Opacity},
  author = {Ikoma, Masahiro and Nakazawa, Kiyoshi and Emori, Hiroyuki},
  year = 2000,
  month = jul,
  journal = {ApJ},
  volume = {537},
  number = {2},
  pages = {1013},
  doi = {10.1086/309050}
}

@article{ImazBlanco2023,
  title = {Inner Edges of Planetesimal Belts: Collisionally Eroded or Truncated?},
  shorttitle = {Inner Edges of Planetesimal Belts},
  author = {Imaz~Blanco, Amaia and Marino, Sebastian and Matr{\`a}, Luca and Booth, Mark and Carpenter, John and Faramaz, Virginie and Henning, Thomas and Hughes, A Meredith and Kennedy, Grant M and P{\'e}rez, Sebasti{\'a}n and Ricci, Luca and Wyatt, Mark C},
  year = 2023,
  month = jul,
  journal = {MNRAS},
  volume = {522},
  number = {4},
  pages = {6150--6169},
  issn = {0035-8711},
  doi = {10.1093/mnras/stad1221},
  urldate = {2025-03-31}
}

@article{Johansen2007,
  title = {Protoplanetary {{Disk Turbulence Driven}} by the {{Streaming Instability}}: {{Nonlinear Saturation}} and {{Particle Concentration}}},
  shorttitle = {Protoplanetary {{Disk Turbulence Driven}} by the {{Streaming Instability}}},
  author = {Johansen, A. and Youdin, A.},
  year = 2007,
  month = jun,
  journal = {ApJ},
  volume = {662},
  number = {1},
  pages = {627},
  publisher = {IOP Publishing},
  issn = {0004-637X},
  doi = {10.1086/516730},
  urldate = {2024-08-16},
  langid = {english}
}

@article{Kanagawa2015,
  title = {Formation of a Disc Gap Induced by a Planet: Effect of the Deviation from {{Keplerian}} Disc Rotation},
  author = {Kanagawa, K. D. and Tanaka, H. and Muto, T. and Tanigawa, T. and Takeuchi, T.},
  year = 2015,
  month = feb,
  journal = {MNRAS},
  volume = {448},
  number = {1},
  eprint = {https://academic.oup.com/mnras/article-pdf/448/1/994/9388300/stv025.pdf},
  pages = {994--1006},
  issn = {0035-8711},
  doi = {10.1093/mnras/stv025}
}

@article{Kanagawa2018,
  title = {Radial Migration of Gap-Opening Planets in Protoplanetary Disks. {{I}}. {{The}} Case of a Single Planet},
  author = {Kanagawa, Kazuhiro D. and Tanaka, Hidekazu and Szuszkiewicz, Ewa},
  year = 2018,
  month = jul,
  journal = {ApJ},
  volume = {861},
  number = {2},
  pages = {140},
  publisher = {The American Astronomical Society},
  doi = {10.3847/1538-4357/aac8d9}
}

@article{Klahr2021,
  title = {Testing the {{Jeans}}, {{Toomre}}, and {{Bonnor}}--{{Ebert Concepts}} for {{Planetesimal Formation}}: {{3D Streaming-instability Simulations}} of {{Diffusion-regulated Formation}} of {{Planetesimals}}},
  shorttitle = {Testing the {{Jeans}}, {{Toomre}}, and {{Bonnor}}--{{Ebert Concepts}} for {{Planetesimal Formation}}},
  author = {Klahr, Hubert and Schreiber, Andreas},
  year = 2021,
  month = apr,
  journal = {ApJ},
  volume = {911},
  number = {1},
  pages = {9},
  publisher = {The American Astronomical Society},
  issn = {0004-637X},
  doi = {10.3847/1538-4357/abca9b},
  urldate = {2024-04-05},
  langid = {english}
}

@article{Kleine2020,
  title = {The {{Non-carbonaceous}}--{{Carbonaceous Meteorite Dichotomy}}},
  author = {Kleine, T. and Budde, G. and Burkhardt, C. and Kruijer, T. S. and Worsham, E. A. and Morbidelli, A. and Nimmo, F.},
  year = 2020,
  month = may,
  journal = {SSRv},
  volume = {216},
  number = {4},
  pages = {55},
  issn = {1572-9672},
  doi = {10.1007/s11214-020-00675-w},
  urldate = {2024-04-04},
  langid = {english}
}

@article{Lambrechts2019,
  title = {Quasi-Static Contraction during Runaway Gas Accretion onto Giant Planets},
  author = {Lambrechts, M. and Lega, E. and Nelson, R. P. and Crida, A. and Morbidelli, A.},
  year = 2019,
  month = oct,
  journal = {A\&A},
  volume = {630},
  pages = {A82},
  publisher = {EDP Sciences},
  issn = {0004-6361, 1432-0746},
  doi = {10.1051/0004-6361/201834413},
  urldate = {2024-04-10},
  copyright = {\copyright{} ESO 2019},
  langid = {english}
}

@article{Lau2022,
  title = {Rapid Formation of Massive Planetary Cores in a Pressure Bump},
  author = {Lau, Tommy Chi Ho and Dr{\k a}{\.z}kowska, Joanna and Stammler, Sebastian M. and Birnstiel, Tilman and Dullemond, Cornelis P.},
  year = 2022,
  journal = {A\&A},
  volume = {668},
  pages = {A170},
  doi = {10.1051/0004-6361/202244864}
}

@article{Lau2023,
  title = {Parallelization of the Symplectic Massive Body Algorithm ({{SyMBA}}) n-Body Code},
  author = {Lau, Tommy Chi Ho and Lee, Man Hoi},
  year = 2023,
  month = apr,
  journal = {RNAAS},
  volume = {7},
  number = {4},
  pages = {74},
  publisher = {The American Astronomical Society},
  doi = {10.3847/2515-5172/accc8a}
}

@article{Lau2024,
  title = {Can the Giant Planets of the {{Solar System}} Form via Pebble Accretion in a Smooth Protoplanetary Disc?},
  author = {Lau, Tommy Chi Ho and Lee, Man Hoi and Brasser, Ramon},
  year = 2024,
  journal = {A\&A},
  volume = {683},
  doi = {10.1051/0004-6361/202347863},
  __markedentry = {[tommy:6]},
  refid = {10.105100046361202347863},
  keywords = {Astrophysics - Earth and Planetary Astrophysics,methods: numerical,planet-disk interactions,planets and satellites: formation},
  file = {/home/tommy/snap/zotero-snap/common/Zotero/storage/6PXAX87D/Lau et al. - 2024 - Can the giant planets of the Solar System form via pebble accretion in a smooth protoplanetary disc.pdf;/home/tommy/snap/zotero-snap/common/Zotero/storage/AA82XKG9/Lau et al. - 2024 - Can the giant planets of the Solar System form via pebble accretion in a smooth protoplanetary disc.pdf}
}

@phdthesis{Lau2024a,
  title = {{Modelling planet formation}},
  author = {Lau, Tommy Chi Ho},
  year = 2024,
  month = nov,
  urldate = {2025-03-20},
  langid = {ngerman},
  school = {Ludwig-Maximilians-Universit\"at M\"unchen}
}

@article{Lau2024b,
  title = {Sequential Giant Planet Formation Initiated by Disc Substructure},
  author = {Lau, Tommy Chi Ho and Birnstiel, Til and Dr{\k a}{\.z}kowska, Joanna and Stammler, Sebastian Markus},
  year = 2024,
  month = aug,
  journal = {A\&A},
  volume = {688},
  pages = {A22},
  publisher = {EDP Sciences},
  issn = {0004-6361, 1432-0746},
  doi = {10.1051/0004-6361/202450464},
  urldate = {2024-07-31},
  copyright = {\copyright{} The Authors 2024},
  langid = {english}
}

@article{Lenz2019,
  title = {Planetesimal Population Synthesis: {{Pebble}} Flux-Regulated Planetesimal Formation},
  author = {Lenz, Christian T. and Klahr, Hubert and Birnstiel, Tilman},
  year = 2019,
  month = mar,
  journal = {ApJ},
  volume = {874},
  number = {1},
  pages = {36},
  publisher = {The American Astronomical Society},
  issn = {0004-637X},
  doi = {10.3847/1538-4357/ab05d9}
}

@article{Lenz2020,
  title = {Constraining the Parameter Space for the Solar Nebula},
  author = {Lenz, Christian T. and Klahr, Hubert and Birnstiel, Tilman and Kretke, Katherine and Stammler, Sebastian},
  year = 2020,
  journal = {A\&A},
  volume = {640},
  doi = {10.1051/0004-6361/202037878},
  __markedentry = {[tommy:]},
  refid = {10.105100046361202037878}
}

@article{Li2021,
  title = {Thresholds for {{Particle Clumping}} by the {{Streaming Instability}}},
  author = {Li, Rixin and Youdin, Andrew N.},
  year = 2021,
  month = sep,
  journal = {ApJ},
  volume = {919},
  number = {2},
  pages = {107},
  publisher = {The American Astronomical Society},
  issn = {0004-637X},
  doi = {10.3847/1538-4357/ac0e9f},
  urldate = {2025-03-21},
  langid = {english}
}

@article{Lim2024,
  title = {Streaming {{Instability}} and {{Turbulence}}: {{Conditions}} for {{Planetesimal Formation}}},
  shorttitle = {Streaming {{Instability}} and {{Turbulence}}},
  author = {Lim, Jeonghoon and Simon, Jacob B. and Li, Rixin and Armitage, Philip J. and Carrera, Daniel and Lyra, Wladimir and Rea, David G. and Yang, Chao-Chin and Youdin, Andrew N.},
  year = 2024,
  month = jul,
  journal = {ApJ},
  volume = {969},
  number = {2},
  pages = {130},
  publisher = {The American Astronomical Society},
  issn = {0004-637X},
  doi = {10.3847/1538-4357/ad47a2},
  urldate = {2025-07-14},
  langid = {english}
}

@article{Lin1986,
  title = {On the {{Tidal Interaction}} between {{Protoplanets}} and the {{Protoplanetary Disk}}. {{III}}. {{Orbital Migration}} of {{Protoplanets}}},
  author = {Lin, D. N. C. and Papaloizou, John},
  year = 1986,
  month = oct,
  journal = {ApJ},
  volume = {309},
  pages = {846},
  publisher = {IOP},
  issn = {0004-637X},
  doi = {10.1086/164653},
  urldate = {2024-05-16}
}

@article{Liu2018,
  title = {Catching Drifting Pebbles - {{I}}. {{Enhanced}} Pebble Accretion Efficiencies for Eccentric Planets},
  author = {Liu, Beibei and Ormel, Chris W.},
  year = 2018,
  journal = {A\&A},
  volume = {615},
  pages = {A138},
  doi = {10.1051/0004-6361/201732307}
}

@article{Liu2022,
  title = {Early {{Solar System}} Instability Triggered by Dispersal of the Gaseous Disk},
  author = {Liu, Beibei and Raymond, Sean N. and Jacobson, Seth A.},
  year = 2022,
  month = apr,
  journal = {Nature},
  volume = {604},
  number = {7907},
  pages = {643--646},
  publisher = {Nature Publishing Group},
  issn = {1476-4687},
  doi = {10.1038/s41586-022-04535-1},
  urldate = {2025-10-25},
  copyright = {2022 The Author(s), under exclusive licence to Springer Nature Limited},
  langid = {english}
}

@article{Lubow2006,
  title = {Gas {{Flow}} across {{Gaps}} in {{Protoplanetary Disks}}},
  author = {Lubow, S. H. and D'Angelo, G.},
  year = 2006,
  month = apr,
  journal = {ApJ},
  volume = {641},
  number = {1},
  pages = {526},
  publisher = {IOP Publishing},
  issn = {0004-637X},
  doi = {10.1086/500356},
  urldate = {2024-08-19},
  langid = {english}
}

@article{Lust1952,
  title = {Die {{Entwicklung}} Einer Um Einen {{Zentralk\"orper}} Rotierenden {{Gasmasse}}. {{I}}. {{L\"osungen}} Der Hydrodynamischen {{Gleichungen}} Mit Turbulenter {{Reibung}}},
  author = {L{\"u}st, R.},
  year = 1952,
  month = jan,
  journal = {ZNatA},
  volume = {7},
  pages = {87--98},
  issn = {0932-0784},
  doi = {10.1515/zna-1952-0118},
  urldate = {2024-04-18}
}

@article{Lynden-Bell1974,
  title = {The {{Evolution}} of {{Viscous Discs}} and the {{Origin}} of the {{Nebular Variables}}},
  author = {{Lynden-Bell}, D. and Pringle, J. E.},
  year = 1974,
  month = sep,
  journal = {MNRAS},
  volume = {168},
  number = {3},
  pages = {603--637},
  issn = {0035-8711},
  doi = {10.1093/mnras/168.3.603},
  urldate = {2024-12-19}
}

@article{MacGregor2017,
  title = {A {{Complete ALMA Map}} of the {{Fomalhaut Debris Disk}}},
  author = {MacGregor, Meredith A. and Matr{\`a}, Luca and Kalas, Paul and Wilner, David J. and Pan, Margaret and Kennedy, Grant M. and Wyatt, Mark C. and Duchene, Gaspard and Hughes, A. Meredith and Rieke, George H. and Clampin, Mark and Fitzgerald, Michael P. and Graham, James R. and Holland, Wayne S. and Pani{\'c}, Olja and Shannon, Andrew and Su, Kate},
  year = 2017,
  month = jun,
  journal = {ApJ},
  volume = {842},
  number = {1},
  pages = {8},
  publisher = {The American Astronomical Society},
  issn = {0004-637X},
  doi = {10.3847/1538-4357/aa71ae},
  urldate = {2025-03-29},
  langid = {english}
}

@article{Marino2017,
  title = {{{ALMA}} Observations of the {$\eta$} {{Corvi}} Debris Disc: Inward Scattering of {{CO-rich}} Exocomets by a Chain of 3--30~{{M}}{$\oplus$} Planets?},
  shorttitle = {{{ALMA}} Observations of the {$\eta$} {{Corvi}} Debris Disc},
  author = {Marino, S. and Wyatt, M. C. and Pani{\'c}, O. and Matr{\`a}, L. and Kennedy, G. M. and Bonsor, A. and Kral, Q. and Dent, W. R. F and Duchene, G. and Wilner, D. and Lisse, C. M. and Lestrade, J.-F. and Matthews, B.},
  year = 2017,
  month = mar,
  journal = {MNRAS},
  volume = {465},
  number = {3},
  pages = {2595--2615},
  issn = {0035-8711},
  doi = {10.1093/mnras/stw2867},
  urldate = {2025-03-29}
}

@article{Marino2021,
  title = {Constraining Planetesimal Stirring: How Sharp Are Debris Disc Edges?},
  shorttitle = {Constraining Planetesimal Stirring},
  author = {Marino, Sebastian},
  year = 2021,
  month = jun,
  journal = {MNRAS},
  volume = {503},
  number = {4},
  pages = {5100--5114},
  issn = {0035-8711},
  doi = {10.1093/mnras/stab771},
  urldate = {2025-03-31}
}

@incollection{Marino2022,
  title = {Planetesimal/{{Debris Disks}}},
  booktitle = {Planetary {{Systems Now}}},
  author = {Marino, Sebastian},
  year = 2022,
  month = may,
  pages = {381--408},
  publisher = {WORLD SCIENTIFIC (EUROPE)},
  doi = {10.1142/9781800613140_0014},
  urldate = {2025-03-29},
  isbn = {978-1-80061-313-3}
}

@article{Mathis1977,
  title = {The Size Distribution of Interstellar Grains.},
  author = {Mathis, J. S. and Rumpl, W. and Nordsieck, K. H.},
  year = 1977,
  month = oct,
  journal = {The Astrophysical Journal},
  volume = {217},
  pages = {425--433},
  publisher = {IOP},
  issn = {0004-637X},
  doi = {10.1086/155591},
  urldate = {2025-03-21}
}

@article{Matra2025,
  title = {{{REsolved ALMA}} and {{SMA Observations}} of {{Nearby Stars}} ({{REASONS}}) - {{A}} Population of 74 Resolved Planetesimal Belts at Millimetre Wavelengths},
  author = {Matr{\`a}, L. and Marino, S. and Wilner, D. J. and Kennedy, G. M. and Booth, M. and Krivov, A. V. and Williams, J. P. and Hughes, A. M. and del Burgo, C. and Carpenter, J. and Davies, C. L. and Ertel, S. and Kral, Q. and Lestrade, J.-F. and Marshall, J. P. and Milli, J. and {\"O}berg, K. I. and Pawellek, N. and Sepulveda, A. G. and Wyatt, M. C. and Matthews, B. C. and MacGregor, M.},
  year = 2025,
  month = jan,
  journal = {A\&A},
  volume = {693},
  pages = {A151},
  publisher = {EDP Sciences},
  issn = {0004-6361, 1432-0746},
  doi = {10.1051/0004-6361/202451397},
  urldate = {2025-03-26},
  copyright = {\copyright{} The Authors 2025},
  langid = {english}
}

@article{Matsumura2017,
  title = {N-Body Simulations of Planet Formation via Pebble Accretion - {{I}}. {{First}} Results},
  author = {Matsumura, Soko and Brasser, Ramon and Ida, Shigeru},
  year = 2017,
  journal = {A\&A},
  volume = {607},
  pages = {A67},
  doi = {10.1051/0004-6361/201731155}
}

@article{Matsumura2021,
  title = {N-Body Simulations of Planet Formation via Pebble Accretion - {{II}}. {{How}} Various Giant Planets Form},
  author = {Matsumura, Soko and Brasser, Ramon and Ida, Shigeru},
  year = 2021,
  month = jun,
  journal = {A\&A},
  volume = {650},
  pages = {A116},
  publisher = {EDP Sciences},
  issn = {0004-6361, 1432-0746},
  doi = {10.1051/0004-6361/202039210},
  urldate = {2025-04-03},
  copyright = {\copyright{} ESO 2021},
  langid = {english}
}

@article{Miller2021,
  title = {The Formation of Wide {{exoKuiper}} Belts from Migrating Dust Traps},
  author = {Miller, E and Marino, S and Stammler, S M and Pinilla, P and Lenz, C and Birnstiel, T and Henning, Th},
  year = 2021,
  month = oct,
  journal = {MNRAS},
  volume = {508},
  number = {4},
  eprint = {https://academic.oup.com/mnras/article-pdf/508/4/5638/41025164/stab2935.pdf},
  pages = {5638--5656},
  issn = {0035-8711},
  doi = {10.1093/mnras/stab2935}
}

@article{Mizuno1980,
  title = {Formation of the Giant Planets},
  author = {Mizuno, Hiroshi},
  year = 1980,
  month = aug,
  journal = {Progress of Theoretical Physics},
  volume = {64},
  number = {2},
  eprint = {https://academic.oup.com/ptp/article-pdf/64/2/544/5289046/64-2-544.pdf},
  pages = {544--557},
  issn = {0033-068X},
  doi = {10.1143/PTP.64.544}
}

@article{Morbidelli2005,
  title = {Chaotic Capture of {{Jupiter}}'s {{Trojan}} Asteroids in the Early {{Solar System}}},
  author = {Morbidelli, A. and Levison, H. F. and Tsiganis, K. and Gomes, R.},
  year = 2005,
  month = may,
  journal = {Nature},
  volume = {435},
  number = {7041},
  pages = {462--465},
  publisher = {Nature Publishing Group},
  issn = {1476-4687},
  doi = {10.1038/nature03540},
  urldate = {2024-12-19},
  copyright = {2005 Macmillan Magazines Ltd.},
  langid = {english}
}

@article{Mousis2017,
  title = {Impact of {{Radiogenic Heating}} on the {{Formation Conditions}} of {{Comet 67P}}/{{Churyumov}}--{{Gerasimenko}}},
  author = {Mousis, O. and Drouard, A. and Vernazza, P. and Lunine, J. I. and Monnereau, M. and Maggiolo, R. and Altwegg, K. and Balsiger, H. and Berthelier, J.-J. and Cessateur, G. and De Keyser, J. and Fuselier, S. A. and Gasc, S. and Korth, A. and Le Deun, T. and Mall, U. and Marty, B. and R{\`e}me, H. and Rubin, M. and Tzou, C.-Y. and Waite, J. H. and Wurz, P.},
  year = 2017,
  month = apr,
  journal = {ApJL},
  volume = {839},
  number = {1},
  pages = {L4},
  publisher = {The American Astronomical Society},
  issn = {2041-8205},
  doi = {10.3847/2041-8213/aa6839},
  urldate = {2025-07-18},
  langid = {english}
}

@article{Nesvorny2018,
  title = {Dynamical {{Evolution}} of the {{Early Solar System}}},
  author = {Nesvorn{\'y}, David},
  year = 2018,
  month = sep,
  journal = {Annual Review of Astronomy and Astrophysics},
  volume = {56},
  number = {Volume 56, 2018},
  pages = {137--174},
  publisher = {Annual Reviews},
  issn = {0066-4146, 1545-4282},
  doi = {10.1146/annurev-astro-081817-052028},
  urldate = {2024-04-24},
  langid = {english}
}

@article{Nesvorny2019,
  title = {Trans-{{Neptunian}} Binaries as Evidence for Planetesimal Formation by the Streaming Instability},
  author = {Nesvorn{\'y}, David and Li, Rixin and Youdin, Andrew N. and Simon, Jacob B. and Grundy, William M.},
  year = 2019,
  month = sep,
  journal = {Nat Astron},
  volume = {3},
  number = {9},
  pages = {808--812},
  publisher = {Nature Publishing Group},
  issn = {2397-3366},
  doi = {10.1038/s41550-019-0806-z},
  urldate = {2025-04-03},
  copyright = {2019 The Author(s), under exclusive licence to Springer Nature Limited},
  langid = {english}
}

@article{Ormel2018,
  title = {Catching Drifting Pebbles - {{II}}. {{A}} Stochastic Equation of Motion for Pebbles},
  author = {Ormel, Chris W and Liu, Beibei},
  year = 2018,
  journal = {A\&A},
  volume = {615},
  pages = {A178},
  doi = {10.1051/0004-6361/201732562}
}

@article{Ormel2021,
  title = {How Planets Grow by Pebble Accretion - {{III}}. {{Emergence}} of an Interior Composition Gradient},
  author = {Ormel, Chris W. and Vazan, Allona and Brouwers, Marc G.},
  year = 2021,
  month = mar,
  journal = {A\&A},
  volume = {647},
  pages = {A175},
  publisher = {EDP Sciences},
  issn = {0004-6361, 1432-0746},
  doi = {10.1051/0004-6361/202039706},
  urldate = {2024-04-10},
  copyright = {\copyright{} ESO 2021},
  langid = {english}
}

@article{Pfeil2024,
  title = {{{TriPoD}}: {{Tri-Population}} Size Distributions for {{Dust}} Evolution - {{Coagulation}} in Vertically Integrated Hydrodynamic Simulations of Protoplanetary Disks},
  shorttitle = {{{TriPoD}}},
  author = {Pfeil, Thomas and Birnstiel, Til and Klahr, Hubert},
  year = 2024,
  month = nov,
  journal = {A\&A},
  volume = {691},
  pages = {A45},
  publisher = {EDP Sciences},
  issn = {0004-6361, 1432-0746},
  doi = {10.1051/0004-6361/202449337},
  urldate = {2025-04-03},
  copyright = {\copyright{} The Authors 2024},
  langid = {english}
}

@article{Picogna2019,
  title = {The Dispersal of Protoplanetary Discs - {{I}}. {{A}} New Generation of {{X-ray}} Photoevaporation Models},
  author = {Picogna, Giovanni and Ercolano, Barbara and Owen, James E. and Weber, Michael L.},
  year = 2019,
  month = jul,
  journal = {MNRAS},
  volume = {487},
  number = {1},
  pages = {691--701},
  issn = {0035-8711}
}

@article{Pinilla2016,
  title = {Can Dead Zones Create Structures like a Transition Disk?},
  author = {Pinilla, Paola and Flock, Mario and Ovelar, Maria de Juan and Birnstiel, Til},
  year = 2016,
  month = dec,
  journal = {A\&A},
  volume = {596},
  pages = {A81},
  publisher = {EDP Sciences},
  issn = {0004-6361, 1432-0746},
  doi = {10.1051/0004-6361/201628441},
  urldate = {2024-04-10},
  copyright = {\copyright{} ESO, 2016},
  langid = {english}
}

@article{Piso2014,
  title = {{{ON THE MINIMUM CORE MASS FOR GIANT PLANET FORMATION AT WIDE SEPARATIONS}}},
  author = {Piso, Ana-Maria A. and Youdin, Andrew N.},
  year = 2014,
  month = apr,
  journal = {ApJ},
  volume = {786},
  number = {1},
  pages = {21},
  publisher = {The American Astronomical Society},
  issn = {0004-637X},
  doi = {10.1088/0004-637X/786/1/21},
  urldate = {2024-12-19},
  langid = {english}
}

@article{Pitjeva2018,
  title = {Mass of the {{Kuiper}} Belt},
  author = {Pitjeva, E. V. and Pitjev, N. P.},
  year = 2018,
  month = sep,
  journal = {Celest Mech Dyn Astr},
  volume = {130},
  number = {9},
  pages = {57},
  issn = {1572-9478},
  doi = {10.1007/s10569-018-9853-5},
  urldate = {2025-05-17},
  langid = {english}
}

@article{Pollack1996,
  title = {Formation of the Giant Planets by Concurrent Accretion of Solids and Gas},
  author = {Pollack, James B. and Hubickyj, Olenka and Bodenheimer, Peter and Lissauer, Jack J. and Podolak, Morris and Greenzweig, Yuval},
  year = 1996,
  journal = {Icarus},
  volume = {124},
  number = {1},
  pages = {62--85},
  issn = {0019-1035},
  doi = {10.1006/icar.1996.0190}
}

@article{Preibisch2005,
  title = {The {{Origin}} of {{T Tauri X-Ray Emission}}: {{New Insights}} from the {{Chandra Orion Ultradeep Project}}},
  shorttitle = {The {{Origin}} of {{T Tauri X-Ray Emission}}},
  author = {Preibisch, Thomas and Kim, Yong-Cheol and Favata, Fabio and Feigelson, Eric D. and Flaccomio, Ettore and Getman, Konstantin and Micela, Giusi and Sciortino, Salvatore and Stassun, Keivan and Stelzer, Beate and Zinnecker, Hans},
  year = 2005,
  month = oct,
  journal = {ApJS},
  volume = {160},
  number = {2},
  pages = {401},
  publisher = {IOP Publishing},
  issn = {0067-0049},
  doi = {10.1086/432891},
  urldate = {2025-03-21},
  langid = {english}
}

@article{Prialnik1987,
  title = {Radiogenic {{Heating}} of {{Comets}} by {{26Al}} and {{Implications}} for {{Their Time}} of {{Formation}}},
  author = {Prialnik, D. and {Bar-Nun}, A. and Podolak, M.},
  year = 1987,
  month = aug,
  journal = {The Astrophysical Journal},
  volume = {319},
  pages = {993},
  publisher = {IOP},
  issn = {0004-637X},
  doi = {10.1086/165516},
  urldate = {2025-07-18}
}

@article{Saito2011,
  title = {{{PLANETESIMAL FORMATION BY SUBLIMATION}}},
  author = {Saito, Etsuko and Sirono, Sin-iti},
  year = 2011,
  month = jan,
  journal = {ApJ},
  volume = {728},
  number = {1},
  pages = {20},
  publisher = {The American Astronomical Society},
  issn = {0004-637X},
  doi = {10.1088/0004-637X/728/1/20},
  urldate = {2024-12-19},
  langid = {english}
}

@article{Schafer2017,
  title = {Initial Mass Function of Planetesimals Formed by the Streaming Instability},
  author = {Sch{\"a}fer, Urs and Yang, Chao-Chin and Johansen, Anders},
  year = 2017,
  month = jan,
  journal = {A\&A},
  volume = {597},
  pages = {A69},
  publisher = {EDP Sciences},
  issn = {0004-6361, 1432-0746},
  doi = {10.1051/0004-6361/201629561},
  urldate = {2025-07-15},
  copyright = {\copyright{} ESO 2017},
  langid = {english}
}

@article{Schoonenberg2018,
  title = {A {{Lagrangian}} Model for Dust Evolution in Protoplanetary Disks: Formation of Wet and Dry Planetesimals at Different Stellar Masses},
  author = {Schoonenberg, Djoeke and Ormel, Chris W. and Krijt, Sebastiaan},
  year = 2018,
  journal = {A\&A},
  volume = {620},
  doi = {10.1051/0004-6361/201834047},
  refid = {10.105100046361201834047}
}

@article{Schreiber2018,
  title = {Azimuthal and {{Vertical Streaming Instability}} at {{High Dust-to-gas Ratios}} and on the {{Scales}} of {{Planetesimal Formation}}},
  author = {Schreiber, Andreas and Klahr, Hubert},
  year = 2018,
  month = jun,
  journal = {ApJ},
  volume = {861},
  number = {1},
  pages = {47},
  publisher = {The American Astronomical Society},
  issn = {0004-637X},
  doi = {10.3847/1538-4357/aac3d4},
  urldate = {2024-12-19},
  langid = {english}
}

@article{Schulik2019,
  title = {Global {{3D}} Radiation-Hydrodynamic Simulations of Gas Accretion: {{Opacity-dependent}} Growth of {{Saturn-mass}} Planets},
  shorttitle = {Global {{3D}} Radiation-Hydrodynamic Simulations of Gas Accretion},
  author = {Schulik, M. and Johansen, A. and Bitsch, B. and Lega, E.},
  year = 2019,
  month = dec,
  journal = {A\&A},
  volume = {632},
  pages = {A118},
  publisher = {EDP Sciences},
  issn = {0004-6361, 1432-0746},
  doi = {10.1051/0004-6361/201935473},
  urldate = {2024-04-10},
  copyright = {\copyright{} ESO 2019},
  langid = {english}
}

@article{Seager2007,
  title = {Mass-{{Radius Relationships}} for {{Solid Exoplanets}}},
  author = {Seager, S. and Kuchner, M. and {Hier-Majumder}, C. A. and Militzer, B.},
  year = 2007,
  month = nov,
  journal = {ApJ},
  volume = {669},
  number = {2},
  pages = {1279},
  publisher = {IOP Publishing},
  issn = {0004-637X},
  doi = {10.1086/521346},
  urldate = {2024-12-19},
  langid = {english}
}

@article{Shakura1973,
  title = {Black Holes in Binary Systems. {{Observational}} Appearance.},
  author = {Shakura, N. I. and Sunyaev, R. A.},
  year = 1973,
  month = jan,
  journal = {A\&A},
  volume = {24},
  pages = {337--355},
  issn = {0004-6361},
  urldate = {2025-03-20}
}

@article{Simon2016,
  title = {{{THE MASS AND SIZE DISTRIBUTION OF PLANETESIMALS FORMED BY THE STREAMING INSTABILITY}}. {{I}}. {{THE ROLE OF SELF-GRAVITY}}},
  author = {Simon, Jacob B. and Armitage, Philip J. and Li, Rixin and Youdin, Andrew N.},
  year = 2016,
  month = may,
  journal = {ApJ},
  volume = {822},
  number = {1},
  pages = {55},
  publisher = {The American Astronomical Society},
  issn = {0004-637X},
  doi = {10.3847/0004-637X/822/1/55},
  urldate = {2025-07-15},
  langid = {english}
}

@incollection{Simon2024,
  title = {Comets and {{Planetesimal Formation}}},
  booktitle = {Comets {{III}}},
  author = {Simon, Jacob B. and Blum, J{\"u}rgen and Birnstiel, Til and Nesvorn{\'y}, David and Dotson, Ren{\'e}e},
  editor = {Meech, {\relax Karen}. J. and Combi, {\relax Michael}. R. and {Bockel{\'e}e-Morvan}, Dominique and Raymond, {\relax Sean}. N. and Zolensky, {\relax Michael}. E.},
  year = 2024,
  eprint = {jj.21819446.9},
  eprinttype = {jstor},
  pages = {63--94},
  publisher = {University of Arizona Press},
  urldate = {2025-03-21},
  isbn = {978-0-8165-5363-1}
}

@article{Squire2020,
  title = {Physical Models of Streaming Instabilities in Protoplanetary Discs},
  author = {Squire, Jonathan and Hopkins, Philip F},
  year = 2020,
  month = sep,
  journal = {MNRAS},
  volume = {498},
  number = {1},
  pages = {1239--1251},
  issn = {0035-8711},
  doi = {10.1093/mnras/staa2311},
  urldate = {2025-04-03}
}

@article{Stammler2022,
  title = {{{DustPy}}: {{A Python Package}} for {{Dust Evolution}} in {{Protoplanetary Disks}}},
  shorttitle = {{{DustPy}}},
  author = {Stammler, Sebastian M. and Birnstiel, Tilman},
  year = 2022,
  month = aug,
  journal = {ApJ},
  volume = {935},
  number = {1},
  pages = {35},
  publisher = {The American Astronomical Society},
  issn = {0004-637X},
  doi = {10.3847/1538-4357/ac7d58},
  urldate = {2024-12-19},
  langid = {english}
}

@article{Stammler2023,
  title = {Leaky Dust Traps: {{How}} Fragmentation Impacts Dust Filtering by Planets},
  shorttitle = {Leaky Dust Traps},
  author = {Stammler, Sebastian Markus and Lichtenberg, Tim and Dr{\k a}{\.z}kowska, Joanna and Birnstiel, Tilman},
  year = 2023,
  month = feb,
  journal = {A\&A},
  volume = {670},
  pages = {L5},
  publisher = {EDP Sciences},
  issn = {0004-6361, 1432-0746},
  doi = {10.1051/0004-6361/202245512},
  urldate = {2024-04-19},
  copyright = {\copyright{} The Authors 2023},
  langid = {english}
}

@article{Szulagyi2016,
  title = {Circumplanetary Disc or Circumplanetary Envelope?},
  author = {Szul{\'a}gyi, J. and Masset, F. and Lega, E. and Crida, A. and Morbidelli, A. and Guillot, T.},
  year = 2016,
  month = aug,
  journal = {MNRAS},
  volume = {460},
  number = {3},
  pages = {2853--2861},
  issn = {0035-8711},
  doi = {10.1093/mnras/stw1160},
  urldate = {2024-04-10}
}

@article{Takahashi2014,
  title = {{{TWO-COMPONENT SECULAR GRAVITATIONAL INSTABILITY IN A PROTOPLANETARY DISK}}: {{A POSSIBLE MECHANISM FOR CREATING RING-LIKE STRUCTURES}}},
  shorttitle = {{{TWO-COMPONENT SECULAR GRAVITATIONAL INSTABILITY IN A PROTOPLANETARY DISK}}},
  author = {Takahashi, Sanemichi Z. and Inutsuka, Shu-ichiro},
  year = 2014,
  month = sep,
  journal = {ApJ},
  volume = {794},
  number = {1},
  pages = {55},
  publisher = {The American Astronomical Society},
  issn = {0004-637X},
  doi = {10.1088/0004-637X/794/1/55},
  urldate = {2024-04-03},
  langid = {english}
}

@article{Tanaka2002,
  title = {Three-{{Dimensional Interaction}} between a {{Planet}} and an {{Isothermal Gaseous Disk}}. {{I}}. {{Corotation}} and {{Lindblad Torques}} and {{Planet Migration}}},
  author = {Tanaka, Hidekazu and Takeuchi, Taku and Ward, William R.},
  year = 2002,
  month = feb,
  journal = {ApJ},
  volume = {565},
  number = {2},
  pages = {1257},
  publisher = {IOP Publishing},
  issn = {0004-637X},
  doi = {10.1086/324713},
  urldate = {2024-12-19},
  langid = {english}
}

@article{Tanigawa2002,
  title = {Gas {{Accretion Flows}} onto {{Giant Protoplanets}}: {{High-Resolution Two-dimensional Simulations}}},
  shorttitle = {Gas {{Accretion Flows}} onto {{Giant Protoplanets}}},
  author = {Tanigawa, Takayuki and Watanabe, Sei-ichiro},
  year = 2002,
  month = nov,
  journal = {ApJ},
  volume = {580},
  number = {1},
  pages = {506},
  issn = {0004-637X},
  doi = {10.1086/343069},
  urldate = {2025-03-22},
  langid = {english}
}

@article{Trilling1998,
  title = {Orbital {{Evolution}} and {{Migration}} of {{Giant Planets}}: {{Modeling Extrasolar Planets}}},
  shorttitle = {Orbital {{Evolution}} and {{Migration}} of {{Giant Planets}}},
  author = {Trilling, D. E. and Benz, W. and Guillot, T. and Lunine, J. I. and Hubbard, W. B. and Burrows, A.},
  year = 1998,
  month = jun,
  journal = {ApJ},
  volume = {500},
  number = {1},
  pages = {428},
  publisher = {IOP Publishing},
  issn = {0004-637X},
  doi = {10.1086/305711},
  urldate = {2025-03-19},
  langid = {english}
}

@article{Tsiganis2005,
  title = {Origin of the Orbital Architecture of the Giant Planets of the {{Solar System}}},
  author = {Tsiganis, K. and Gomes, R. and Morbidelli, A. and Levison, H. F.},
  year = 2005,
  month = may,
  journal = {Nature},
  volume = {435},
  number = {7041},
  pages = {459--461},
  publisher = {Nature Publishing Group},
  issn = {1476-4687},
  doi = {10.1038/nature03539},
  urldate = {2024-12-19},
  copyright = {2005 Macmillan Magazines Ltd.},
  langid = {english}
}

@article{Wada2013,
  title = {Growth Efficiency of Dust Aggregates through Collisions with High Mass Ratios},
  author = {Wada, Koji and Tanaka, Hidekazu and Okuzumi, Satoshi and Kobayashi, Hiroshi and Suyama, Toru and Kimura, Hiroshi and Yamamoto, Tetsuo},
  year = 2013,
  month = nov,
  journal = {A\&A},
  volume = {559},
  pages = {A62},
  publisher = {EDP Sciences},
  issn = {0004-6361, 1432-0746},
  doi = {10.1051/0004-6361/201322259},
  urldate = {2024-08-16},
  copyright = {\copyright{} ESO, 2013},
  langid = {english}
}

@article{Weidenschilling1977,
  title = {Aerodynamics of Solid Bodies in the Solar Nebula},
  author = {Weidenschilling, S. J.},
  year = 1977,
  month = sep,
  journal = {MNRAS},
  volume = {180},
  number = {2},
  eprint = {https://academic.oup.com/mnras/article-pdf/180/2/57/18325207/mnras180-0057.pdf},
  pages = {57--70},
  issn = {0035-8711},
  doi = {10.1093/mnras/180.2.57}
}

@inproceedings{Whipple1972,
  title = {On Certain Aerodynamic Processes for Asteroids and Comets},
  booktitle = {Plasma {{Planet}}},
  author = {Whipple, F. L.},
  editor = {Elvius, Aina},
  year = 1972,
  month = jan,
  pages = {211}
}

@article{Wurm2005,
  title = {Growth of Planetesimals by Impacts at {$\sim$}25 m/s},
  author = {Wurm, Gerhard and Paraskov, Georgi and Krauss, Oliver},
  year = 2005,
  month = nov,
  journal = {Icarus},
  volume = {178},
  number = {1},
  pages = {253--263},
  issn = {0019-1035},
  doi = {10.1016/j.icarus.2005.04.002},
  urldate = {2024-08-16}
}

@article{Yang2017,
  title = {Concentrating Small Particles in Protoplanetary Disks through the Streaming Instability},
  author = {Yang, C.-C. and Johansen, A. and Carrera, D.},
  year = 2017,
  month = oct,
  journal = {A\&A},
  volume = {606},
  pages = {A80},
  publisher = {EDP Sciences},
  issn = {0004-6361, 1432-0746},
  doi = {10.1051/0004-6361/201630106},
  urldate = {2025-03-21},
  copyright = {\copyright{} ESO, 2017},
  langid = {english}
}

@article{Youdin2005,
  title = {Streaming Instabilities in Protoplanetary Disks},
  author = {Youdin, Andrew N. and Goodman, Jeremy},
  year = 2005,
  month = feb,
  journal = {ApJ},
  volume = {620},
  number = {1},
  pages = {459},
  doi = {10.1086/426895}
}
\bibliographystyle{aasjournal}

\appendix
\section{Gas drag and planet-disk interactions} \label{sec:m:mig}
The treatments for aerodynamic gas drag and planet-disk interactions were identical to those presented in \cite{Lau2024} and are summarized below.

We applied the aerodynamic gas drag prescription by \cite{Adachi1976}, which is
\begin{equation}
	\mbox{\boldmath $a$}_\mathrm{drag}=-\left( \frac{3C_\mathrm{D}\rho(z_\mathrm{p})}{8R_\mathrm{p}\rho_\mathrm{s}}\right) v_\mathrm{rel}\mbox{\boldmath $v$}_\mathrm{rel}
\end{equation}
with the distance from the midplane $z_\mathrm{p}$, the drag coefficient $C_\mathrm{D}$ and, the relative velocity between the particle and the gas $\mbox{\boldmath $v$}_\mathrm{rel}$. The gas flow was assumed to be laminar and cylindrical with $|\mbox{\boldmath $v$}_\mathrm{rel}| = r\Omega_\mathrm{K}(1-|\eta|)$. As the planetesimals were in and beyond the kilometer-scale, the large Reynolds number case was generally applicable, where $C_\mathrm{D}=0.5$ \citep{Whipple1972}.

For damping and migration in the low-mass regime, we applied the prescription based on dynamical friction by \cite{Ida2020}. The timescales for the isothermal case and finite $i$, while $i<\hat{h}_\mathrm{g}$, (Appendix D of \cite{Ida2020}) were implemented. The evolution timescales of semimajor axis, eccentricity, and inclination are defined, respectively, by
\begin{equation}
	\tau_a\equiv-\frac{a}{\mathrm{d}a/\mathrm{d}t},\tau_e\equiv-\frac{e}{\mathrm{d}e/\mathrm{d}t} \ \mathrm{and} \ \tau_i\equiv-\frac{i}{\mathrm{d}i/\mathrm{d}t}.
\end{equation}
With $\hat{e}\equiv e/\hat{h}_\mathrm{g}$ and $\hat{i}\equiv i/\hat{h}_\mathrm{g}$, these timescales are given, respectively, by
\begin{equation}
	\tau_a = \frac{t_\mathrm{wav}}{C_\mathrm{T}\hat{h}_\mathrm{g}^2}\left[ 1+\frac{C_\mathrm{T}}{C_\mathrm{M}}\sqrt{\hat{e}^2+\hat{i}^2}\right], \label{eq:tau_a}
\end{equation}
\begin{equation}\label{eq:tau_e}
	\tau_e =1.282t_\mathrm{wav}\left[ 1+ \frac{(\hat{e}^2+\hat{i}^2)^{3/2}}{15} \right]
\end{equation}
and
\begin{equation}
	\tau_i =1.838t_\mathrm{wav}\left[ 1+ \frac{(\hat{e}^2+\hat{i}^2)^{3/2}}{21.5} \right].
\end{equation}
The characteristic time $t_\mathrm{wav}$ \citep{Tanaka2002} is given by
\begin{equation}
	t_\mathrm{wav}=\left( \frac{M_\odot}{m}\right) \left( \frac{M_\odot}{\Sigma_{\mathrm{g}}r^2}\right) \left( \frac{\hat{h}_\mathrm{g}^4}{\Omega_\mathrm{K}}\right).\label{eq:twav}
\end{equation}
With $p_\Sigma\equiv-\mathrm{d}\ln\Sigma_{\mathrm{g}}/\mathrm{d}\ln r$ and $q_T\equiv-\mathrm{d}\ln T/\mathrm{d}\ln r$, the normalized torques $C_\mathrm{M}$ and $C_\mathrm{T}$ are given by
\begin{equation}
	C_\mathrm{M}=6(2p_\Sigma-q_T+2)  \label{C_M}
\end{equation}
and
\begin{equation}
	C_\mathrm{T}=2.73+1.08p_\Sigma+0.87q_T.\label{C_T}
\end{equation}
The three evolution timescales were then applied with the equation of motion, which is
\begin{equation}
	\mbox{\boldmath $a$}=-\frac{v_\mathrm{K}}{2\tau_a}\mbox{\boldmath $e$}_\theta-\frac{v_r}{\tau_e}\mbox{\boldmath $e$}_r-\frac{v_\theta-v_\mathrm{K}}{\tau_e}\mbox{\boldmath $e$}_\theta-\frac{v_z}{\tau_i}\mbox{\boldmath $e$}_z
\end{equation}
in the cylindrical coordinates $(r,\theta,z)$ with the velocity of the planet $\mbox{\boldmath $v$}=(v_r,v_\theta,v_z)$. The local Keplerian velocity $v_\mathrm{K}$ was evaluated at the instantaneous $r$ of the particle to avoid numerical issue as noted in \cite{Lau2022}.

As a planet grows and opens a gap in the disk, \cite{Kanagawa2018} suggested that the magnitude of the torque scales linearly with the gas surface density. This leads to a smooth transition to the high-mass regime of planet migration. Since the dependence of the migration strength on $\Sigma_{\mathrm{g}}$ is already included in the above implementation, this transition was captured in our model when combined with planetary gap opening (Sect. \ref{sec:m:gap}).

\end{CJK*}
\end{document}